\documentclass[12pt,preprint]{aastex}
\usepackage{graphicx}
\usepackage{natbib}

\shorttitle{Gravitational Lensing by the supermassive black hole
in M31} \shortauthors{Bozza, Calchi Novati \& Mancini}

\begin{document}

\title{Gravitational lensing by the supermassive black hole in the center of M31}

\author{V. Bozza$^{1,2}$, S. Calchi Novati$^{1,2}$ and L. Mancini$^{1,2}$}

\affil{$^1$Dipartimento di Fisica ``E.R. Caianiello'',
Universit\`a di Salerno, via S. Allende, Baronissi (SA), Italy.}

\affil{$^2$Istituto Nazionale di Fisica Nucleare, Sezione di
Napoli, Italy.}

\begin{abstract}
We examine the possibility of observing gravitational lensing in
the weak deflection regime by the supermassive black hole in the
center of the galaxy M31. This black hole is significantly more
massive than the black hole in the center of our Galaxy qualifying
itself as a more effective lens. However, it is also more distant
and the candidate stellar sources appear consequently fainter. As
potential sources we separately consider stars belonging to the
bulge, to the disk, to the triple nucleus formed by P1+P2 and by
the recently discovered inner cluster P3. We calculate the number
of simultaneously lensed stars at a given time as a function of
the threshold magnitude required for the secondary image. For
observations in the K-band we find 1.4 expected stars having
secondary images brighter than $K=24$ and 182 brighter than
$K=30$. For observations in the $V$-band we expect 1.3 secondary
images brighter than $V=27$ and 271 brighter than $V=33$. The
bulge stars have the highest chance to be lensed by the
supermassive black hole, whereas the disk and the composite
nucleus stars contribute by 10$\%$ each. The typical angular
separation of the secondary images from the black hole range from
1 mas to $0.1''$. For each population we also show the
distribution of the lensed sources as a function of their distance
and absolute magnitude, the expected angular positions and
velocities of the generated secondary images, the rate and the
typical duration of the lensing events.
\end{abstract}


\section{Introduction}

As soon as the evidence of a supermassive black hole in the center
of the Milky Way (identified with the radio source Sgr A*) became
overwhelming, the scientific community started to investigate its
ability of acting as a gravitational lens for background stellar
sources. The first work by \citet{WarYus} considered lensing of
sources in the immediate environment of Sgr A* and showed that the
background sky should appear slightly depleted by a lensing
effect. They suggested the possibility that gravitational lensing
could be strong enough to generate an observable secondary image
for several stars at a given time. More precise estimates by
\citet{Jar}, \citet{AleSte}, and \citet{CGM}, point out that about
10 stars belonging to the Galactic bulge should give rise to
secondary images with $K<23$ at a given time. Deep and high
resolution images in the near infrared of the Galactic center are
being obtained by several advanced observatories such as Keck,
VLT, and Gemini North. The progress registered by these
observations encourages to search for signatures of lensing
effects by Sgr A*. For example, identifying the two images of a
background source \citep{Ale}, or measuring the astrometric shift
of the background stars due to gravitational lensing
\citep{NusBro}, could help to determine the position of Sgr A*
very accurately. Enhanced microlensing caused by black holes
surrounding Sgr A* has also been considered \citep{AleLoe,CGM}. A
fascinating possibility is offered by stars orbiting very close to
Sgr A*, which have now been followed very accurately across
several years
\citep{EckGen,Genzel,Ghez1,Eckart,Sch1,Sch2,Ghez2,Ghez3,Weinberg,Eis,Paumard,Reid}.
In fact, the precise knowledge of the position of the source
allows to predict the time, the position and the brightness of the
secondary image \citep{DeP,BozMan1,BozMan2}.

Meanwhile, our knowledge about the central regions of the
Andromeda galaxy (M31) has grown considerably. \citet{Kent} has
provided a detailed brightness and density profile for the bulge
and the disk of M31, which is still the reference for present
works. \citet{Lauer93} discovered that the nucleus of M31 is
actually constituted by two components, which they called P1 and
P2. \citet{King95} noticed that P2 is much brighter than P1 in the
ultraviolet. Then \citet{Tremaine} hypothesized that P1 and P2 are
actually parts of the same eccentric disk orbiting a supermassive
black hole, with P1 being formed by stars at the apocenter and P2
being formed by stars at the pericenter. Later on,
\citet{Bender05} discovered a bright cluster of young stars (named
P3) embedded within P2. They also managed to give the most precise
estimate for the mass of the central black hole of M31,
$M=1.4^{+0.9}_{-0.3}\times 10^8$ M$_\sun$. Recently,
\citet{DemVir} have cast doubt on the hypothesis that the P3
cluster is made of young stars of spectral class A5 -- B5,
proposing that it is actually made of old stars. On the other
hand, \citet{CMCQ} have supported the young-stars hypothesis
suggesting a mechanism for the refuelling of gas in the
neighborhood of the central black hole.

It is very interesting to note that the supermassive black hole in
the center of M31 is nearly two orders of magnitudes heavier than
the black hole in our Galaxy, estimated to $3.6\times 10^6$
M$_\sun$ by \citet{Eis}. From the lensing point of view, the
greater distance of M31 with respect to Sgr A* is therefore partly
compensated by the larger mass of its central black hole. This
motivates a deep investigation of the probability of having
stellar sources lensed by the supermassive black hole in M31.

Gravitational lensing effects have been successfully employed in
several contexts as a tool to study the structure of galaxies,
large-scale structures and cosmological parameters
\citep{SaasFee}. As we will show in this paper, it is not unlikely
that gravitational lensing by the supermassive black hole in M31
will follow the wake of its ancestors. With a conspicuous number
of events it will be possible to undertake a precise
reconstruction of the mass distribution in the inner core of M31.
Gravitational lensing would then provide an independent and
unbiased method to be crossed with other investigation methods
such as object counting, spectroscopic or proper motion
measurements and so on. Indeed the study of the physics of the
core of M31 would greatly benefit from these new data, which will
help to understand the physics of the stellar environment of
supermassive black holes and possiblly shed light on the true
origin of these enigmatic objects.

In this paper, we shall exploit the present knowledge of the bulge
and the disk populations in M31 to calculate the expected number
of simultaneously lensed sources for a given threshold magnitude
in either the $K$-band or the $V$-band. We will also consider
sources belonging to the central clusters, modelling P1 and P2 as
components of a single eccentric system and P3 as a separate inner
stellar cluster surrounding the central black hole. For each of
the four populations (disk, bulge, P1+P2, P3), we shall present
several probability distributions characterizing the properties of
the lensing events. The paper is structured as follows: in \S~2 we
indicate our reference models for the source populations and
describe their features. In \S~3 we review some basics on
gravitational lensing, with particular reference to the black hole
in M31. In \S~4 we present our estimates for the number of lensed
sources at any given time for a given threshold magnitude of the
secondary image. In \S~5 we examine the contribution of sources of
different magnitude to the total number of events. In \S~6 we give
the distribution of events as a function of the source distance.
In \S~7 we show the distribution of the angular positions of the
images. In \S~8 we present the distribution of their apparent
angular velocities. In \S~9 we estimate the rate and the average
duration of the lensing events. In \S~10 we discuss some issues
concerning the identification of the lensing events. \S~11
contains the conclusions.

\section{Source populations} \label{Sec Source}

Our aim is to calculate lensing probabilities for four different
populations of stellar sources around the supermassive black hole
in the center of M31, namely P3, P1+P2, the bulge and the disk. We
therefore introduce specific models for each population containing
information on their spatial distribution, kinematic properties
and luminosity function, which are the necessary elements for the
calculation of the lensing probabilities to be treated in the
succeeding sections.

For all populations, we choose to normalize the spatial
distribution $f_{P_j}$ to unity, so that the quantity
$f_{P_j}(x,y,z)dxdydz$ represents the probability of finding a
single star of the population $P_j$ in the space element $dxdydz$.

We shall present gravitational lensing probability estimates for
hypothetic observation programs in the $V$-band ($\sim$ 0.55
$\mu$m) or in the $K$-band ($\sim$ 2.2 $\mu$m). The $K$-band
presents some considerable advantages for this kind of research,
since the interstellar extinction is lower in the near infrared
than in the visible. In fact, in the $K$-band the total average
extinction (M31 + our Galaxy) is $A_K=0.1$ \citep{Olsen06},
whereas in the $V$-band we have $A_V=0.31$. The latter value can
be deduced by adding the intrinsic extinction of the M31 galaxy
$A_{V,\mathrm{int}}=0.12$ (half the value given by \citealt{Han},
since our sources are close to the center of M31) to the
foreground extinction $A_{V,\mathrm{ext}}=0.19$ \citep{Schlegel}.
Moreover, large ground-based telescopes are optimized for
interferometry in the infrared bands. In the most powerful
configuration, they can reach resolutions of order the mas. Such
capability would be very precious for the detection of secondary
images generated in gravitational lensing events. Even the
forthcoming {\it James Webb Space Telescope (JWST)} is designed to
carry instruments for deep infrared imaging \citep{Gar}. On the
other hand, the $V$-band is less affected by background noise.
Furthermore, at the diffraction limit, the $V$-band enjoys a
resolution 4 times better than the $K$-band, though large
interferometers operating in the visible bands are still far to
come.

In any case, by comparing the results in the two bands we get a
much deeper understanding on the source selection operated by the
gravitational lensing phenomenon. Since the stellar populations
have different luminosity functions in the two bands, all
gravitational lensing distributions look different, with some
features exalted or depressed. This also allows a quick
double-check of our results.

On the basis of this choice, we need the luminosity functions of
each source population both in the $K$-band and in the $V$-band.
In the $K$-band, this function will be expressed as
$n_{P_j}(M_K)$, defined so that $n_{P_j}(M_K)dM_K$ represents the
number of stars belonging to the population $P_j$ with absolute
magnitude in the range $[M_K,M_K+dM_K]$. The total number of stars
belonging to the population $P_j$ is recovered after integration
on all magnitudes
\begin{equation}
N^\mathrm{tot}_{P_j}=\int n_{P_j}(M_K)dM_K.
\end{equation}

The same definitions hold for the $V$-band, with the obvious
changes in the notation. The following subsections explain the
construction of the luminosity functions and the choice of the
spatial distribution for each stellar population.

\subsection{P3}

The cluster P3 was discovered by \citet{Bender05} through
spectroscopic observations using the {\it Hubble Space Telescope
(HST)}. Although this cluster is embedded within P2 (see
\S~\ref{Sec P2}), it is characterized by a distinct stellar
population and different kinematic properties, strongly indicating
that it must be considered as a separate entity.

The characteristics of P3 are consistent with the hypothesis of a
circular disk of stars in Keplerian rotation around the central
supermassive black hole. In particular, \citet{Bender05} used two
models to fit the spectroscopic observations: an exponential flat
disk and a \citet{Sch79} triaxial model. The best fits were
obtained with the flat disk or a thin Schwarzschild model, with
axial ratio 0.26. As a reasonable synthesis of the models explored
by \citet{Bender05}, we choose a classical thick-disk spatial
distribution
\begin{equation}
f_{\mathrm{P3}}(x,y,z)=\frac{1}{4\pi
r_\mathrm{P3}^2z_\mathrm{P3}}\exp\left[-\frac{\sqrt{x^2+y^2}}{r_\mathrm{P3}}\right]
\mathrm{sech}^2\left[\frac{z}{z_\mathrm{P3}}\right]
\end{equation}
with the disk scale being $r_\mathrm{P3}=0.8$ pc and thickness
$z_\mathrm{P3}=0.1$ pc. The inclination of the disk relative to
the line of sight is $i_\mathrm{P3}=55^\circ$, similar to that of
the system P1+P2.

The rotation curve of P3 is symmetric around its center, reaching
a rotation velocity $v_\mathrm{P3}=618 \mathrm{~ km ~ s}^{-1}$ and
dispersion $\sigma_\mathrm{P3}=674 \mathrm{~ km ~ s}^{-1}$.

\begin{figure}
\begin{center}
\resizebox{8cm}{!}{\includegraphics{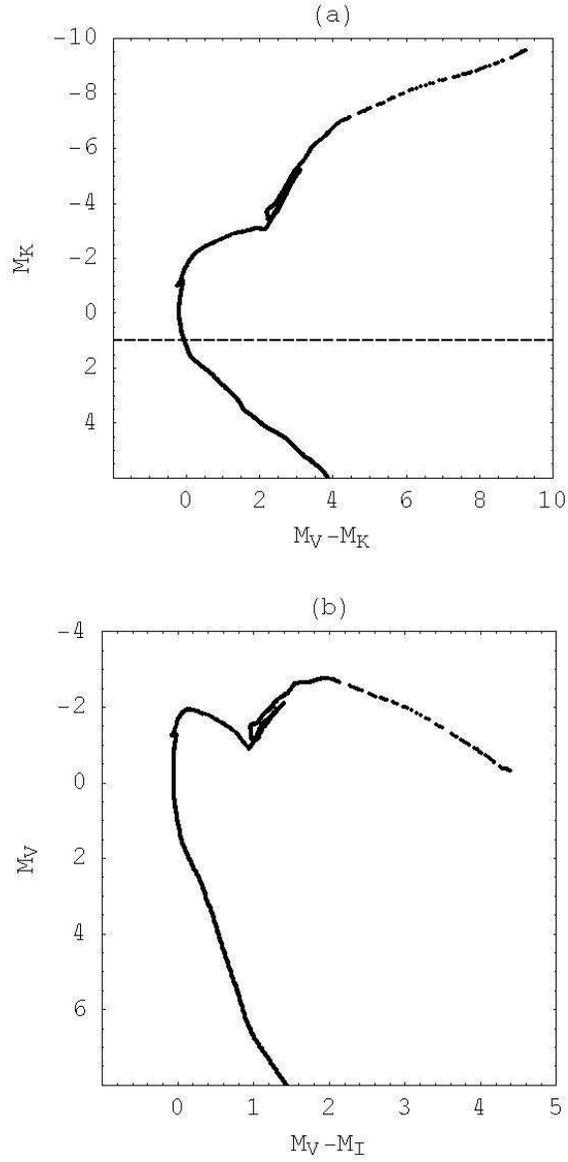}}
 \caption{(a) ($M_K,M_V-M_K$) color-magnitude diagram for P3; the dashed line represents the
 boundary
 between the two samples of stars re-combined with different weights in the luminosity
 function. (b) ($M_V,M_V-M_I$) color-magnitude diagram for P3.
}
 \label{Fig LumP3}
 \end{center}
\end{figure}

For the luminosity function of P3 we have followed
\citet{Bender05}, who use the synthetic color-magnitude diagram
generated by the program IAC-STAR by \citet{ApaGal}. We have run
this program with the same parameters, accepting the hypothesis
that P3 was generated by a single starburst that occurred 200 Myr
ago in a gas cloud with solar metallicity (see \citealt{DemVir}
for an alternative proposal). We adopt a \citet{KTG} initial mass
function. The algorithm has returned us $10^5$ stars with $M_K<6$.
Such a number turns out to be insufficient for a substantial
covering of the giant stars branches. We have thus generated a
second sample of $10^5$ stars with $M_K<1$ and combined the two
samples with appropriate weights. The two samples are shown
together in two different color-magnitude diagrams in Figure
\ref{Fig LumP3}. The cut-off at $M_K=6$ does not translate into a
sharp cut-off in the $V$-band. However, the sample is completely
unaffected for $M_V< 8$, which we take as our cut-off in the
luminosity function in the $V$-band. It is important to stress
that gravitational-lensing observations select very bright sources
with much higher probability than faint ones. For this reason, we
do not pay much attention to the completeness of our simulated
samples of stars on the low-luminosity side. In the same way,
issues concerning the lower cutoff of the initial mass function
are not relevant for us. All gravitational lensing probabilities
are practically insensitive to changes in the source distribution
function at low luminosities. This statement is also supported
{\it a posteriori} by the distributions presented in \S~\ref{Sec
Mag}.

Note that the red giant branch (RGB) lies at almost fixed
magnitude in the $V$-band, whereas it spans 6 magnitudes in the
$K$-band up to the tip.

The normalization of our sample of P3 stars has been performed
introducing a factor $N$ multiplying the total number of stars.
The value of $N$ has been deduced by comparing the total magnitude
in the $V$-band of our artificial set with the total magnitude of
P3 as deduced by \citet{Bender05}, being
$M_{V,{\mathrm{P3}}}=-5.7$.
\begin{equation}
M_{V,{\mathrm{P3}}}=-2.5\log_{10}N-2.5\log_{10}M_{V,{\mathrm{P3-sample}}},
\end{equation}
where $M_{V,{\mathrm{P3-sample}}}$ is the unnormalized absolute
magnitude of our combined sample. Also this normalization
procedure on the number of stars deduced by the total luminosity
is largely insensitive to the abundance of low-luminosity stars.

Finally, the same set of stars has been used to build a binned
luminosity function in the $K$-band and in the $V$-band,
normalized in the way just described. The values of these binned
functions are shown in Tables \ref{Tab LumK} and \ref{Tab LumV},
together with those of the other source populations, to be
discussed in the succeeding sections.

The different orientation of the RGB in the $V$-band and in the
$K$-band is reflected in the luminosity function, which is zero in
the $V$-band up to $M_V\simeq-2.75$, whereas it is already
non-zero in the $K$-band at $M_K\simeq-9.75$. The $K$-band
luminosity function clearly shows a red clump at $M_K\simeq
-3.75$, which is not as clear in the $V$-band, where it overlaps
the stars at the turn-off point (TOP) of the main sequence (MS).
It is worth noting that the real number of stars in P3 should be
of order a few thousands, whereas we have simulated $2\times 10^5$
stars in order to have a statistically significant sample.

\begin{table}
\centering

\begin{tabular}{ccccc}
 \hline \hline

  $M_K$ & $n_{\mathrm{P3}}$ & $n_{\mathrm{P1+P2}}\times 10^{-3}$& $n_{\mathrm{Bulge}}\times 10^{-6}$& $n_{\mathrm{Disk}}\times 10^{-6}$ \\
  \hline
-10.25 & 0 & 0 & 0 & 0.0054 \\
-9.75 & 0.06 & 0 & 0 & 0.0054 \\
-9.25 & 0.038 & 0 & 0 & 0.032 \\
-8.75 & 0.04 & 0 & 0 & 0.07 \\
-8.25 & 0.02 & 0 & 0 & 0.12 \\
-7.75 & 0.024 & 0 & 0 & 0.12 \\
-7.25 & 0.038 & 0.12 & 0.15 & 0.2 \\
-6.75 & 0.07 & 0.54 & 0.68 & 0.98 \\
-6.25 & 0.095 & 0.86 & 1.1 & 1.5 \\
-5.75 & 0.097 & 0.9 & 1.1 & 1.5 \\
-5.25 & 0.49 & 1.1 & 1.4 & 2. \\
-4.75 & 1.1 & 1.6 & 2. & 2.9 \\
-4.25 & 1.3 & 2.1 & 2.6 & 4.2 \\
-3.75 & 11 & 3.5 & 4.4 & 6.8 \\
-3.25 & 3.6 & 5.5 & 6.9 & 12 \\
-2.75 & 0.091 & 5.2 & 6.5 & 12 \\
-2.25 & 0.17 & 7.9 & 9.8 & 26 \\
-1.75 & 0.18 & 34 & 42 & 124 \\
-1.25 & 7.4 & 42 & 53 & 35 \\
-0.75 & 17 & 26 & 33 & 38 \\
-0.25 & 29 & 24 & 30 & 47 \\
0.25 & 49 & 33 & 41 & 81 \\
0.75 & 90 & 48 & 60 & 152 \\
1.25 & 127 & 67 & 83 & 297 \\
1.75 & 237 & 121 & 151 & 527 \\
2.25 & 254 & 349 & 436 & 910 \\
2.75 & 300 & 677 & 846 & 1550 \\
3.25 & 309 & 1317 & 1645 & 2249 \\
3.75 & 443 & 2272 & 2839 & 3375 \\
4.25 & 555 & 3505 & 4380 & 4613 \\
4.75 & 511 & 4027 & 5032 & 4436 \\
5.25 & 752 & 4600 & 5748 & 6587 \\
5.75 & 899 & 6754 & 8440 & 8272 \\
 \hline

\end{tabular}
\caption{Binned luminosity functions in the $K$-band for P3,
P1+P2, the bulge and the disk of M31. As each bin spans 0.5
magnitudes, the number of stars in each bin is just the tabulated
value of $n_{P_j}$ multiplied by 0.5.}\label{Tab LumK}
\end{table}

\begin{table}
\centering

\begin{tabular}{ccccc}
 \hline \hline
  $M_V$ & $n_{\mathrm{P3}}$ & $n_{\mathrm{P1+P2}}\times 10^{-3}$& $n_{\mathrm{Bulge}}\times 10^{-6}$& $n_{\mathrm{Disk}}\times 10^{-6}$ \\
  \hline
-8.25 & 0 & 0 & 0 & 0.0054 \\
-7.75 & 0 & 0 & 0 & 0 \\
-7.25 & 0 & 0 & 0 & 0.027 \\
-6.75 & 0 & 0 & 0 & 0.0054 \\
-6.25 & 0 & 0 & 0 & 0.0054 \\
-5.75 & 0 & 0 & 0 & 0 \\
-5.25 & 0 & 0 & 0 & 0.022 \\
-4.75 & 0 & 0 & 0 & 0.038 \\
-4.25 & 0 & 0 & 0 & 0.054 \\
-3.75 & 0 & 0 & 0 & 0.12 \\
-3.25 & 0 & 0 & 0 & 0.25 \\
-2.75 & 0.22 & 0.0052 & 0.0065 & 0.37 \\
-2.25 & 0.74 & 0.062 & 0.078 & 0.84 \\
-1.75 & 7. & 0.32 & 0.4 & 1.7 \\
-1.25 & 21 & 0.96 & 1.2 & 4.1 \\
-0.75 & 22 & 2.5 & 3.1 & 12 \\
-0.25 & 35 & 6.5 & 8.1 & 30 \\
0.25 & 53 & 9.7 & 12 & 57 \\
0.75 & 80 & 51 & 63 & 165 \\
1.25 & 98 & 31 & 38 & 97 \\
1.75 & 118 & 26 & 33 & 148 \\
2.25 & 138 & 33 & 41 & 213 \\
2.75 & 132 & 41 & 51 & 329 \\
3.25 & 150 & 64 & 80 & 551 \\
3.75 & 166 & 235 & 294 & 928 \\
4.25 & 178 & 826 & 1033 & 1382 \\
4.75 & 205 & 1021 & 1276 & 1542 \\
5.25 & 223 & 1317 & 1646 & 1748 \\
5.75 & 233 & 1512 & 1890 & 1932 \\
6.25 & 247 & 1706 & 2131 & 2046 \\
6.75 & 248 & 1754 & 2192 & 2073 \\
7.25 & 258 & 1897 & 2370 & 2227 \\
7.75 & 299 & 2253 & 2815 & 2609 \\
 \hline
\end{tabular}
\caption{Binned luminosity functions in the $V$-band for P3,
P1+P2, the bulge and the disk of M31.}\label{Tab LumV}
\end{table}

\subsection{P1+P2} \label{Sec P2}

On a scale a bit larger than P3, the supermassive black hole in
M31 is surrounded by two larger clusters, called P1 and P2.
Actually, P1 and P2 are parts of the same eccentric disk orbiting
the central black hole, with P2 being composed of stars at the
periapsis and P1 by stars at the apoapsis \citep{Tremaine,PeiTre}.

\begin{figure}
\begin{center}
\resizebox{8cm}{!}{\includegraphics{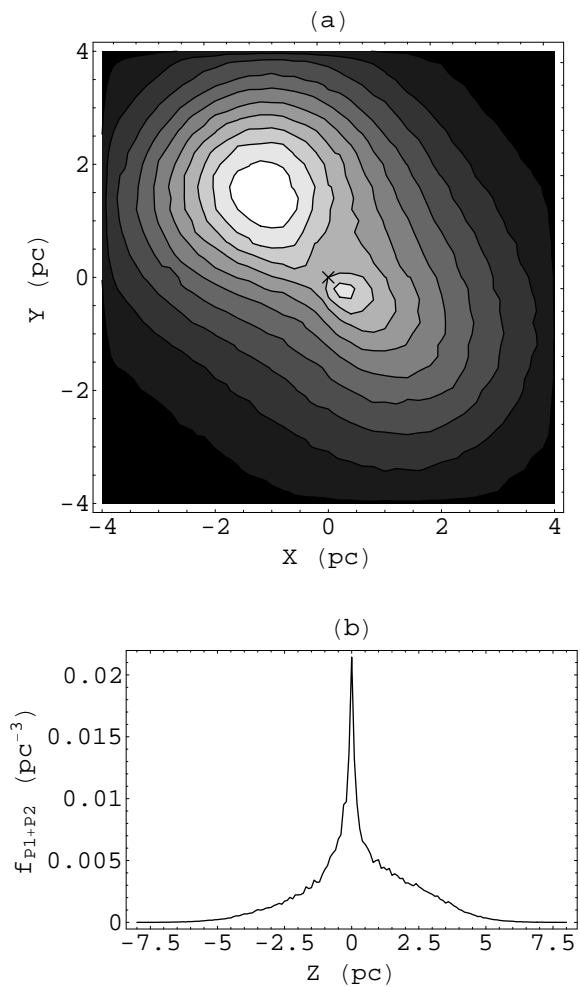}}
 \caption{(a) Reconstructed brightness profile of the central
 cluster P1+P2, obtained following the prescriptions of
 \citet{PeiTre}; north is up and east is left;
 the cross indicates the position of the central black hole. P1 is
 the brighter cluster, whereas P2 is the smaller one appearing
 closer to the black hole.
 (b) Spatial distribution of the cluster P1+P2 along the line connecting
 the observer with the black hole. The positive $Z$-axis points toward the observer.}
 \label{Fig P1+P2}
 \end{center}
\end{figure}

The spatial distribution of the whole system was obtained by
\citet{PeiTre} fitting the brightness profile of a simulated
sample of $10^7$ stars with orbital parameters randomly chosen
from some suitable distributions. We have followed their steps,
simulating the same number of stars using the best fit parameters
of their non-aligned model, which provides an excellent fit to the
observed brightness profile of P1+P2. In Figure \ref{Fig P1+P2}a
we show the brightness profile thus obtained. For all the details
relative to the eccentric disk model, the reader is referred to
\citet{PeiTre}. To give an idea of the shape of the cluster P1+P2,
we just mention that the scale of its extension is fixed by the
parameter $a_0=1.37$ pc, its inclination along the line of sight
is $\theta_i=54.1^\circ$, its thickness is roughly controlled by
the combination $a_0\sigma_I^0$, with $\sigma_I^0=24.6^\circ$. For
every star, the eccentricity is a function of the semiaxis of the
orbit with several parameters. It is roughly peaked at $e=0.5$.

Binning our simulated distribution of stars, we then easily obtain
the spatial distribution function $f_{P1+P2}(x,y,z)$, to be used
in the lensing calculations.

In the kinematic calculations, since the stars behind the black
hole are closer to P2, we have used the average rotation velocity
of P2, estimated as $v_\mathrm{P2}=220 \mathrm{~ km ~ s}^{-1}$
with dispersion $\sigma_\mathrm{P2}=100 \mathrm{~ km ~ s}^{-1}$
\citep{PeiTre}.

The characteristics of the stars composing P2 and P1 are quite
similar to those of typical bulge stars. We have therefore used
the bulge sample illustrated in \S~\ref{Sec Bulge}. Here we just
mention that we have normalized this sample in the $V$-band, using
$V_{\mathrm{P1+P2}}=12.55$ mag \citep{PeiTre}. With this
normalization, we have built the luminosity functions in the
$K$-band and in the $V$-band, as shown in Table \ref{Tab LumK}.
Some considerations on these functions are included in the
following section.

\subsection{Bulge} \label{Sec Bulge}

The bulge of M31 has been studied in great detail by \citet{Kent},
who traced precise luminosity and density contours. By
interpolating these contours, it is possible to build a very
accurate spatial distribution. We shall refer to this distribution
as $f_{\mathrm{Bulge}}$. The inclination of the plane of symmetry
of the bulge is $i_\mathrm{Bulge}=77^\circ$. The bulge is assumed
to have negligible rotation and constant dispersion velocity
$\sigma_\mathrm{Bulge}=160 \mathrm{~ km ~ s}^{-1}$.

\begin{figure}
\begin{center}
\resizebox{8cm}{!}{\includegraphics{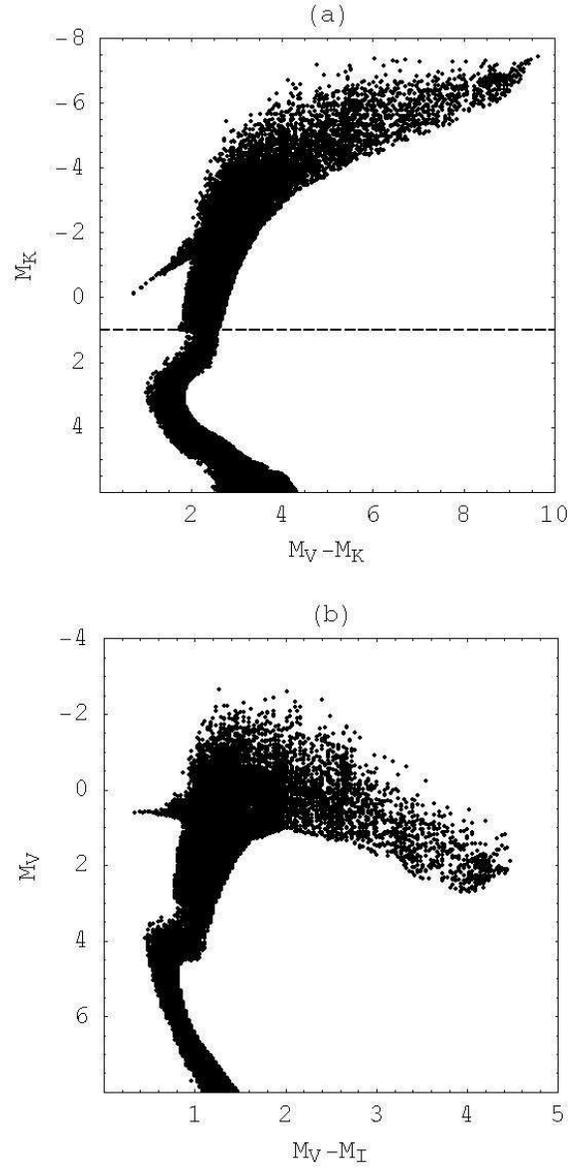}}
 \caption{(a) ($M_K,M_V-M_K$) color-magnitude diagram for the bulge and the cluster P1+P2;
 the dashed line represents the boundary
 between the two samples of stars re-combined with different weights in the luminosity
 function. (b) ($M_V,M_V-M_I$) color-magnitude diagram for the bulge and P1+P2.
}
 \label{Fig LumBulge}
 \end{center}
\end{figure}

In order to generate viable luminosity functions for the bulge, we
have used the program IAC-STAR by \citet{ApaGal} in this case as
well. In particular, following \citet{SarJab}, we have supposed
that the bulge (and also P1+P2) were generated by a single
starburst occurred 12.6 Gyr ago with metallicity following a
closed box law
\begin{equation}
\frac{dN}{dZ}=\frac{1}{y}e^{(Z-Z_0)/y},
\end{equation}
with $Z_0=0$ and yield $y$ equal to the solar metallicity $Z_\sun
=0.019$. With these specifications, we have generated $10^5$ stars
with $M_K<6$ and $10^5$ stars with $M_K<1$. Before combining the
two samples, we have randomly removed stars at the lower end of
the metallicity distribution ($Z<0.004$), in order to reproduce
the data of \citet{SarJab} as accurately as possible. This
additional cut has reduced the two samples by roughly 10\%. Figure
\ref{Fig LumBulge} shows two color-magnitude diagrams of the
combined sample of bulge stars. By comparing with the real ones
studied by \citet{SarJab}, it is possible to appreciate the
accuracy of our sample.

The luminosity profile introduced by \citet{Kent} is obtained in
the $r$-band of the Thuan and Gunn filter set. Assuming $r-K=2.9$
\citep{Olsen06}, we can directly normalize the luminosity function
in the $K$-band and consequently in the $V$-band. The final
results are included in Tables \ref{Tab LumK} and \ref{Tab LumV}.

The bulge RGB lies in the range $-1>M_K>-7.5$. The TOP is at
$M_K\simeq -3$ rather than $M_K=0$ as in P3. The red clump is
peaked at $M_K=-1.25$ and is also evident in the $V$-band at
$M_V=0.75$.

\subsection{Disk} \label{Sec Disk}

The disk spatial distribution is modelled by
\begin{equation}
f_{\mathrm{Disk}}(x,y,z)=\frac{1}{4\pi
r_\mathrm{Disk}^2z_\mathrm{Disk}}\exp\left[-\frac{\sqrt{x^2+y^2}}{r_\mathrm{Disk}}\right]
\mathrm{sech}^2\left[\frac{z}{z_\mathrm{Disk}}\right],
\end{equation}
with disk scale $r_\mathrm{Disk}=5.5$ kpc, thickness
$z_\mathrm{Disk}=0.3$ kpc and inclination
$i_\mathrm{Disk}=77^\circ$ \citep{WidDub}.

The rotation velocity of the stars in the disk is
$v_\mathrm{Disk}=250 \mathrm{~ km ~ s}^{-1}$ with a negligible
dispersion \citep{Kent}.

The disk star formation history is quite different from the one
followed by the bulge and presumably by the central cluster stars.
Following \citet{Hodge}, \citet{Wil} and \citet{Bel03}, we take a
simple Population I model with a constant star formation rate
throughout the history of M31 and constant solar metallicity. With
these assumptions, we have run the program IAC-STAR obtaining
$10^5$ stars with $M_K<6$ and more $10^5$ stars with $M_K<1$. The
color-magnitude diagrams so-obtained are shown in Figure \ref{Fig
LumDisk}.

\begin{figure}
\begin{center}
\resizebox{8cm}{!}{\includegraphics{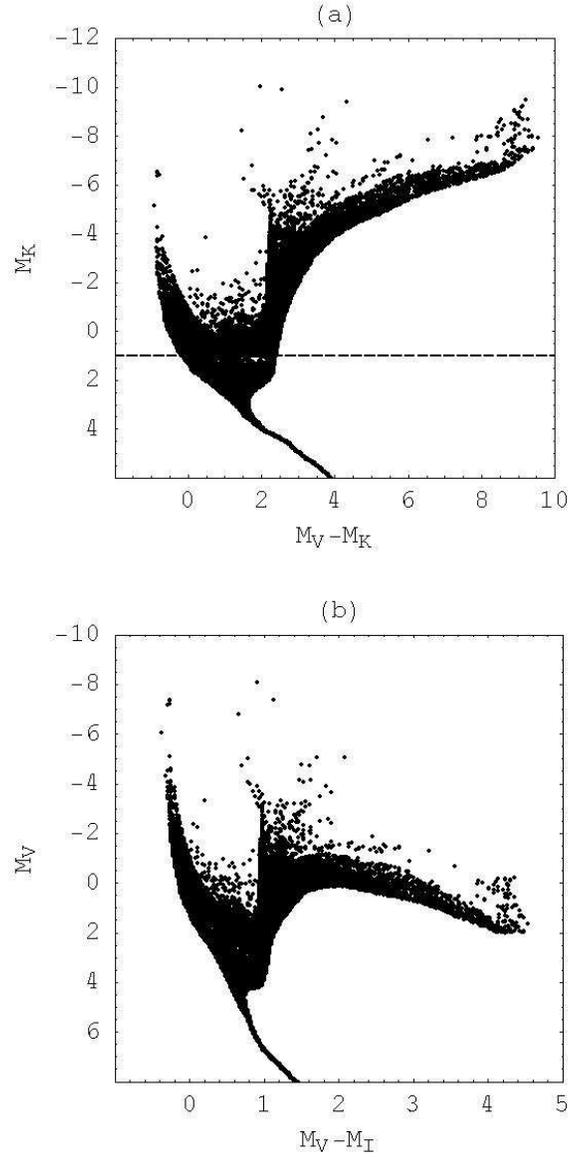}}
 \caption{(a) ($M_K,M_V-M_K$) color-magnitude diagram for the disk;
 the dashed line represents the boundary
 between the two samples of stars re-combined with different weights in the luminosity
 function. (b) ($M_V,M_V-M_I$) color-magnitude diagram for the disk.
}
 \label{Fig LumDisk}
 \end{center}
\end{figure}

The normalization of our sample of stars has been determined
comparing the total luminosity in the $R$-band to the total
luminosity as indicated by \citet{WidDub}
$R_{\mathrm{Disk}}=-21.4$. Finally, the correctly normalized
luminosity function in the $K$-band and in the $V$-band have been
included in Tables \ref{Tab LumK} and \ref{Tab LumV}.

With respect to the other populations, the disk has a complete
main sequence extending up to $M_K\simeq -7$ and $M_V\simeq -7$.
Moreover, there are also some Red Supergiants (RSG) at $M_K=-10$
and $M_V=-8$. The existence of these supergiants mainly affects
the $V$-band luminosity function, since it makes the disk
luminosity function start at much brighter magnitudes with respect
to the other populations. The red clump is peaked at $M_K=-1.75$
and $M_V=0.75$, though almost contiguous to the MS in the
$V$-band.

\section{Basics of gravitational lensing}

In this section we shall briefly recall some definitions used in
standard gravitational lensing. This will prepare the ground for
the analysis to be presented in the following sections.

The supermassive black hole in the center of M31 can be modelled
as a simple point lens, with mass $M=1.4\times 10^8$ M$_\sun$,
placed at a distance $D_\mathrm{OL}=760$ kpc from the Sun
\citep{Bender05}. We define the optical axis as the line joining
the observer to the lens. For a source behind the black hole at
distance $D_\mathrm{LS}$ from it, the lens equation takes the form
\begin{equation}
\beta=\theta-\frac{\theta_{\mathrm{E}}^2}{\theta}, \label{LensEq}
\end{equation}
where $\beta$ is the angle between the line joining the observer
to the source and the optical axis, $\theta$ is the angle formed
by the observed image with the optical axis and
\begin{equation}
\theta_{\mathrm{E}}=\sqrt{\frac{4GM}{c^2}\frac{D_\mathrm{LS}}{D_\mathrm{OL}D_\mathrm{OS}}}
\label{ThetaEin}
\end{equation}
is the Einstein angle and $D_\mathrm{OS}=D_\mathrm{OL} +
D_\mathrm{LS}$ is the distance from the observer to the source. As
the source distances considered in this paper are at most of the
order of a few kpc behind the lens, we generally have
$D_\mathrm{OS}\simeq D_\mathrm{OL}$.

Solving the lens equation for $\theta$, we obtain the position of
the two images
\begin{equation}
\theta_\pm = \frac{1}{2}\left( \beta \pm \sqrt{\beta^2
+4\theta_\mathrm{E}^2} \right).
\end{equation}

These two images are magnified by a factor
\begin{equation}
\mu_\pm=\frac{u^2+2}{2u \sqrt{u^2+4}} \pm \frac{1}{2}, \label{mu}
\end{equation}
where $u=\beta/\theta_\mathrm{E}$ is the source angular position
normalized to the Einstein angle.

\begin{figure}[t]
\begin{center}
\resizebox{10cm}{!}{\includegraphics{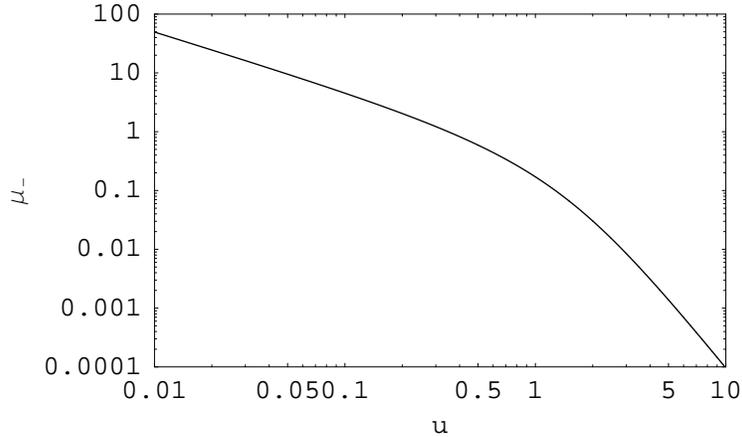}}
 \caption{Magnification of the secondary image as a function of the source position normalized to
 the Einstein angle.
}
 \label{Fig DiskRad}
 \end{center}
\end{figure}

Figure \ref{Fig DiskRad} shows the magnification of the secondary
image as a function of the normalized source position. It clearly
shows how the dependence of the magnification on $u$ changes from
$u^{-1}$ to $u^{-4}$ as we go from the regime of good alignment
($u \ll 1$) to the regime of bad alignment ($u \gg 1$), the
Einstein angle representing the scale of the transition between
the two regimes.

When the source is far from the optical axis ($\beta \gg
\theta_\mathrm{E}$) the secondary image has very low magnification
and becomes unobservable. Our aim is to determine the probability
of seeing the secondary images of sources in M31, which will
typically become observable when $\beta$ is of the same order of
magnitude as $\theta_\mathrm{E}$. We then note that the angular
separation of the secondary image from the central black hole is
always of the order of the Einstein angle for $\beta \lesssim
\theta_\mathrm{E}$. The order of magnitude of the Einstein angle
in the physical situation we are considering is
\begin{equation}
\theta_\mathrm{E}=44\, \mathrm{mas} \left(
\frac{D_\mathrm{LS}}{\mathrm{kpc}} \right)^{1/2}.
\end{equation}

The Einstein angle is of order $0.1''$ for sources as far as a few
kpc, whereas it drops to a few mas when the source is closer than
100 pc. We can thus conclude that in order to observe the
secondary images of sources closer than 1 kpc to the central black
hole we need very high angular resolution, only achievable by
interferometric techniques, such as those employed in the VLTI,
Keck and LBT. This gives an advantage to source populations
distributed on a larger scale (bulge and disk) with respect to the
central clusters P1+P2 and P3. However, distinguishing a secondary
image at $0.1''$ from the central black hole from other sources is
by no means easy and certainly demands high resolution as well.

Since we are speaking about lensing by a supermassive black hole,
it is interesting to check that the weak deflection paradigm holds
for all the interesting events. Higher order images formed by
photons performing loops around the central black hole show up
very close to the black hole shadow border, which has a radius of
$\theta_\mathrm{sh}=9$ $\mu$as \citep{Dar,VirEll,Boz1}. In order
to detect such images, a much greater effort with respect to
ordinary lensing images would be necessary. For this reason, we
will not consider them in the present analysis.

It remains to check whether the secondary image can always be
correctly described in the weak deflection limit. This can be
verified by comparing the position of the image $\theta_-$ with
the radius of the shadow border $\theta_\mathrm{sh}$.
Approximating $\theta_-$ by $\theta_\mathrm{E}$, the ratio
$\theta_\mathrm{sh}/\theta_\mathrm{E}$ gives an estimate of the
error we commit by neglecting the next to leading order term in
the deflection angle \citep{KeePet}. It is easy to calculate that
the error is $2\%$ at $D_\mathrm{LS}=0.1$ pc and $6\%$ at
$D_\mathrm{LS}=0.01$ pc. As it will be evident in \S~\ref{Sec
Dis}, where we calculate the distribution of the events as a
function of the source distance, only the low $D_\mathrm{LS}$ tail
of the P3 distribution is affected by these errors and then only
marginally. Considering the low relevance of this tail in the
total estimates and the uncertainties in the P3-population
modelling, we will simply ignore any strong deflection effect in
our analysis.

\section{Number of lensed sources}

This section contains the main calculation of this work, namely
the number of expected lensing events at a given time for each
source population around the supermassive black hole in M31. We
first introduce the methodology that we have followed and then
present our estimates for each population at the end of the
section. We will focus on the $K$-band, with analogous
considerations holding for the $V$-band.

Consider a source of absolute magnitude $M_K$. The observed
magnitude $K$ is
\begin{equation}
K=M_K+5\log_{10}\frac{D_{\mathrm{OS}}}{10\mathrm{~ pc}}+ A_K,
\label{Ko}
\end{equation}
where $A_K=0.1$ is the extinction in the $K$-band \citep{Olsen06}.
If the source suffers gravitational lensing, its secondary image
has apparent magnitude
\begin{equation}
K_{-}= K -2.5\log_{10}\mu_-, \label{K-}
\end{equation}
with $\mu_-$ given by equation (\ref{mu}).

If we fix a threshold magnitude $K_{\mathrm{thr}}$ for the
detection of the secondary image, only the sources sufficiently
magnified will have a secondary image with $K_-<K_{\mathrm{thr}}$.
The minimum magnification needed to bring a source with absolute
magnitude $M_K$ above threshold can be found by simply inverting
equation (\ref{K-}) with $K_{-}=K_\mathrm{thr}$
\begin{equation}
\mu_-=10^{-0.4(\Delta K-24.5)},
\end{equation}
where $\Delta K = K_\mathrm{thr} - M_K$.

As the magnification is a function of the normalized source
position angle $u$ through equation (\ref{mu}), there exists a
limiting value for $u$ (which we shall indicate by $u_\mathrm{Z}$)
such that a source with absolute magnitude $M_K$ has a secondary
image just at the threshold value $K_{\mathrm{thr}}$. This can be
found inverting equation (\ref{mu})
\begin{equation}
u_\mathrm{Z} (\mu_-)= \left[ 2
\mu_{-}(1+\mu_{-})+(1+2\mu_{-})\sqrt{ \mu_{-}(1+\mu_{-})}
\right]^{-1/2}.
\end{equation}

Defining $\beta_\mathrm{Z}= \theta_\mathrm{E} u_\mathrm{Z}$, all
sources with $\beta<\beta_\mathrm{Z}$ have a secondary image
brighter than $K_{\mathrm{thr}}$. Finally, we can also define the
radius of the lensing zone at distance $D_\mathrm{LS}$, as the
radius of the circle containing the sources with magnitude $M_K$
whose secondary image is magnified above $K_{\mathrm{thr}}$. Of
course, the radius of this circle is simply
$R_\mathrm{Z}=\beta_{\mathrm{Z}} D_\mathrm{OS}$. For any value of
the distance $D_\mathrm{LS}$, only sources with $r<R_\mathrm{Z}$
have an observable secondary image. As $D_{\mathrm{LS}}$ varies,
we can thus define a lensing zone with radius
$R_\mathrm{Z}(D_{\mathrm{LS}})$ centered on the optical axis and
that contains all the sources with absolute magnitude $M_K$ that
give rise to observable gravitational lensing effects. More
explicitly, the radius of the lensing zone is
\begin{equation}
R_\mathrm{Z} = 0.16 \, \mathrm{pc} \left(
\frac{D_\mathrm{LS}}{\mathrm{kpc}} \right)^{1/2}
u_\mathrm{Z}\left(\mu_-(\Delta K) \right),
\end{equation}
which clearly shows that the radius of the lensing zone grows with
the square root of the distance of the source from the black hole.
The dependence on the absolute magnitude of the source and the
threshold fixed by observations is stored in the function
$u_\mathrm{Z}\left(\mu_-(\Delta K) \right)$. The brighter the
source and the fainter the threshold, the larger the lensing zone.

The probability of finding a single star of absolute magnitude
$M_K$ in the lensing zone defined by the threshold magnitude
$K_{\mathrm{thr}}$ is obtained integrating the spatial
distribution of the population to which the source belongs in the
domain contained within the lensing zone
\begin{equation}
\Pi_{P_j}(M_K,K_{\mathrm{thr}})=\int\limits_0^\infty d
D_\mathrm{LS} \int\limits_0^{R_{\mathrm{Z}}} dr ~ r
\int\limits_0^{2\pi} d\phi f_{P_j}(D_\mathrm{LS},r,\phi),
\label{PiPj}
\end{equation}
where we recall that $R_\mathrm{Z}$ is a function of
$D_\mathrm{LS}$, $M_K$, and $K_\mathrm{thr}$. In the evaluation of
the spatial distribution one must take care of the correct
geometric orientation in the space of the population considered,
as specified in \S~\ref{Sec Source}.

Finally, the total number of lensed sources with a secondary above
threshold is
\begin{equation}
N_{P_j}(K_{\mathrm{thr}})= \int n_{P_j}(M_K)
\Pi_{P_j}(M_K,K_{\mathrm{thr}}) \; dM_K. \label{NPj}
\end{equation}

\begin{figure}
\begin{center}
\resizebox{10cm}{!}{\includegraphics{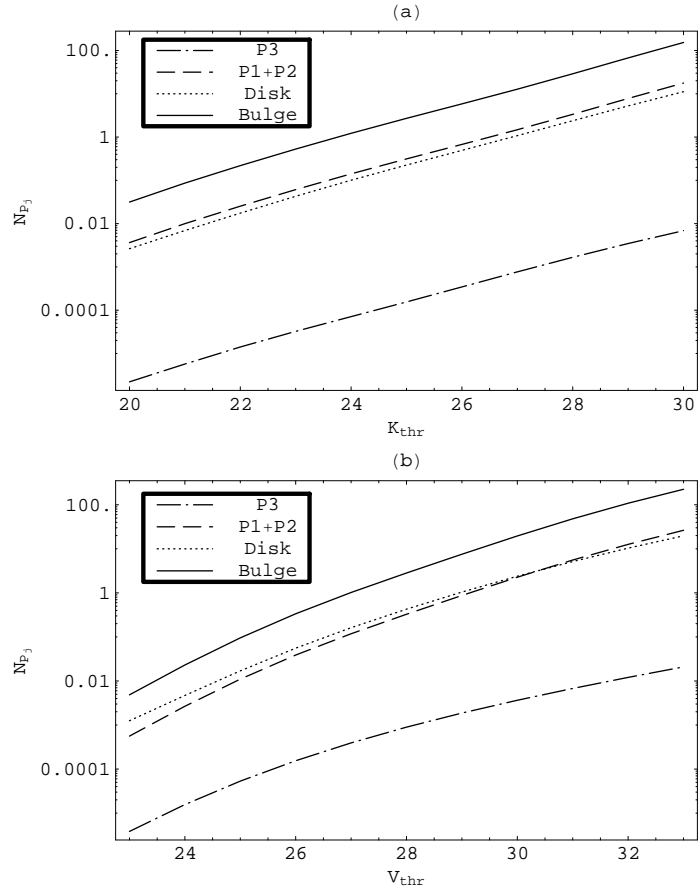}}
 \caption{(a) Number of expected lensing events as a function of
 the threshold magnitude for the secondary image in the $K$-band.
 (b) Number of expected lensing events as a function of
 the threshold magnitude for the secondary image in the $V$-band.
 Note the different range in the ordinate axis in the two plots.
}
 \label{Fig NEvents}
 \end{center}
\end{figure}

Figure \ref{Fig NEvents}a shows the estimated number of lensing
events for each of the four source populations considered in this
paper as a function of the threshold magnitude $K_\mathrm{thr}$.
Repeating the same steps, we can get a similar plot for
observations lead in the $V$-band, with a threshold magnitude
$V_\mathrm{thr}$ as shown in Figure \ref{Fig NEvents}b.

We see that P3 gives a practically negligible contribution. The
stars in P3 are too few and too close to the supermassive black
hole. The lensing zone is indeed too restricted at distances of
the order of a pc to get a sizeable number of events.

The situation for P1+P2 is much better, because this cluster
extends to larger radii and is more populated. However, we have to
push the threshold magnitude to $K_\mathrm{thr}=27$ or
$V_\mathrm{thr}=30$ in order to have at least one expected event.
The estimated number of events for the disk is very similar to
that of P1+P2.

The bulge is by far the best reservoir of good sources for
gravitational lensing by the central black hole. The number of
expected events is already larger than one at $K_\mathrm{thr}=24$
or $V_\mathrm{thr}=27$, reaching more than one hundred at
$K_\mathrm{thr}=30$ or $V_\mathrm{thr}=32$. Comparing the plots in
the $K$-band and the $V$-band, we can note that redder populations
such as P1+P2 and the bulge are slightly depressed when going from
the $K$-band to the $V$-band. The disk supergiants, not present in
other populations, keep the the disk number of events higher at
low values of $V_\mathrm{thr}$ with respect to the other
populations.

Roughly, the plot in the $V$-band is very similar to the plot in
the $K$-band, but is shifted to higher values of $V_\mathrm{thr}$
by 2.5 magnitudes. This apparent gap between the two bands is
actually completely recovered when one compares the signal to
noise ratio (SNR) in the two bands. In fact, a source with $V-K=0$
emits 5.77 photons in the $V$-band for each photon in the $K$-band
\citep{5.77}. In the background-limited regime, the expected noise
due to the 4 populations considered in this paper can be easily
estimated thanks to the spatial distributions and luminosity
functions introduced in \S ~ \ref{Sec Source}. Indeed we find that
for images with the same FWHM the SNR in the $V$-band is 2.5
magnitudes better than in the $K$-band. This justifies our choice
to plot our expectations in the intervals $20<K_\mathrm{thr}<30$
and $23<V_\mathrm{thr}<33$. Furthermore, if the diffraction limit
in the two bands is reached with the same aperture, one gets an
additional bonus of 1.5 magnitudes for the $V$-band. However, at
the present time, the largest interferometers such as VLTI, Keck
and LBT are not designed for observations in the $V$-band and
therefore only the $K$-band can take advantage of the resolutions
available at such long-baseline facilities.

The expectations plotted in Figure \ref{Fig NEvents} are also
summarized in Tables \ref{Tab NEventsK} and \ref{Tab NEventsV}.
Recall that these estimates give the number of secondary images
above threshold simultaneously present at a given time. See
\S~\ref{Sec Vel} and \S~\ref{Sec Kin} for a discussion on their
evolution in time. It should be noted that the estimates presented
in this section do not take into account the angular separation of
the secondary images from the central black hole. \S~\ref{Sec Ima}
is devoted to that issue.

\begin{table}
\centering

\begin{tabular}{cccc}
 \hline \hline

  $K_\mathrm{thr}$ & $N_{\mathrm{Bulge}}$ & $N_{\mathrm{Disk}}$ &  $N_{\mathrm{P1+P2}}$ \\
  \hline
20 & 0.032 & 0.0026 & 0.0036 \\
21 & 0.086 & 0.0069 & 0.01 \\
22 & 0.22 & 0.017 & 0.025 \\
23 & 0.53 & 0.043 & 0.061 \\
24 & 1.2 & 0.1 & 0.14 \\
25 & 2.7 & 0.23 & 0.31 \\
26 & 5.8 & 0.49 & 0.68 \\
27 & 13 & 1.1 & 1.5 \\
28 & 29 & 2.4 & 3.4 \\
29 & 67 & 5.3 & 7.8 \\
30 & 153 & 11 & 18 \\
 \hline

\end{tabular}
\caption{Expected number of events for sources in the bulge, the
disk and P1+P2 for different threshold magnitudes in the $K$-band
for the secondary image.}\label{Tab NEventsK}
\end{table}

\begin{table}
\centering
\begin{tabular}{cccc}
 \hline \hline

  $V_\mathrm{thr}$ & $N_{\mathrm{Bulge}}$ & $N_{\mathrm{Disk}}$ &  $N_{\mathrm{P1+P2}}$ \\
  \hline
23 & 0.0048 & 0.0013 & 0.00056 \\
24 & 0.023 & 0.0047 & 0.0027 \\
25 & 0.095 & 0.017 & 0.011 \\
26 & 0.34 & 0.056 & 0.039 \\
27 & 1. & 0.16 & 0.12 \\
28 & 2.8 & 0.43 & 0.33 \\
29 & 7.5 & 1. & 0.87 \\
30 & 20 & 2.4 & 2.3 \\
31 & 48 & 5.1 & 5.6 \\
32 & 108 & 10 & 13 \\
33 & 225 & 20 & 26 \\
 \hline

\end{tabular}
\caption{Expected number of events for sources in the bulge, the
disk and P1+P2 for different threshold magnitudes in the $V$-band
for the secondary image.}\label{Tab NEventsV}
\end{table}

\section{Magnitude of the lensed sources} \label{Sec Mag}

\begin{figure*}
\begin{center}
\resizebox{\hsize}{!}{\includegraphics{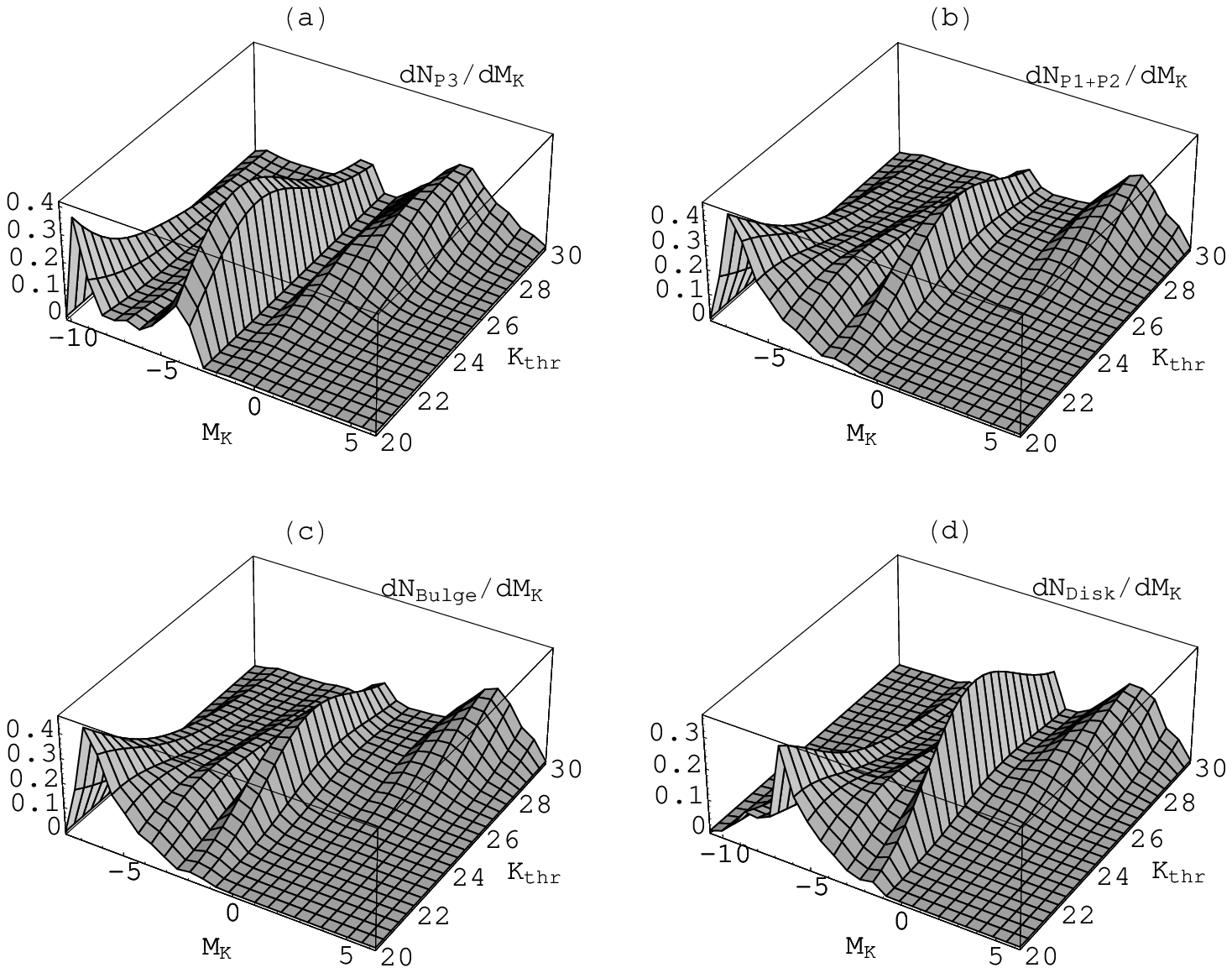}}
 \caption{Relative contribution of sources with different absolute
 magnitudes to the total number of expected events for threshold magnitudes ranging from 20 to 30 in the
 $K$-band. (a) P3; (b) P1+P2; (c) bulge; (d) disk.
}
 \label{Fig dNdK}
 \end{center}
\end{figure*}

\begin{figure*}
\begin{center}
\resizebox{\hsize}{!}{\includegraphics{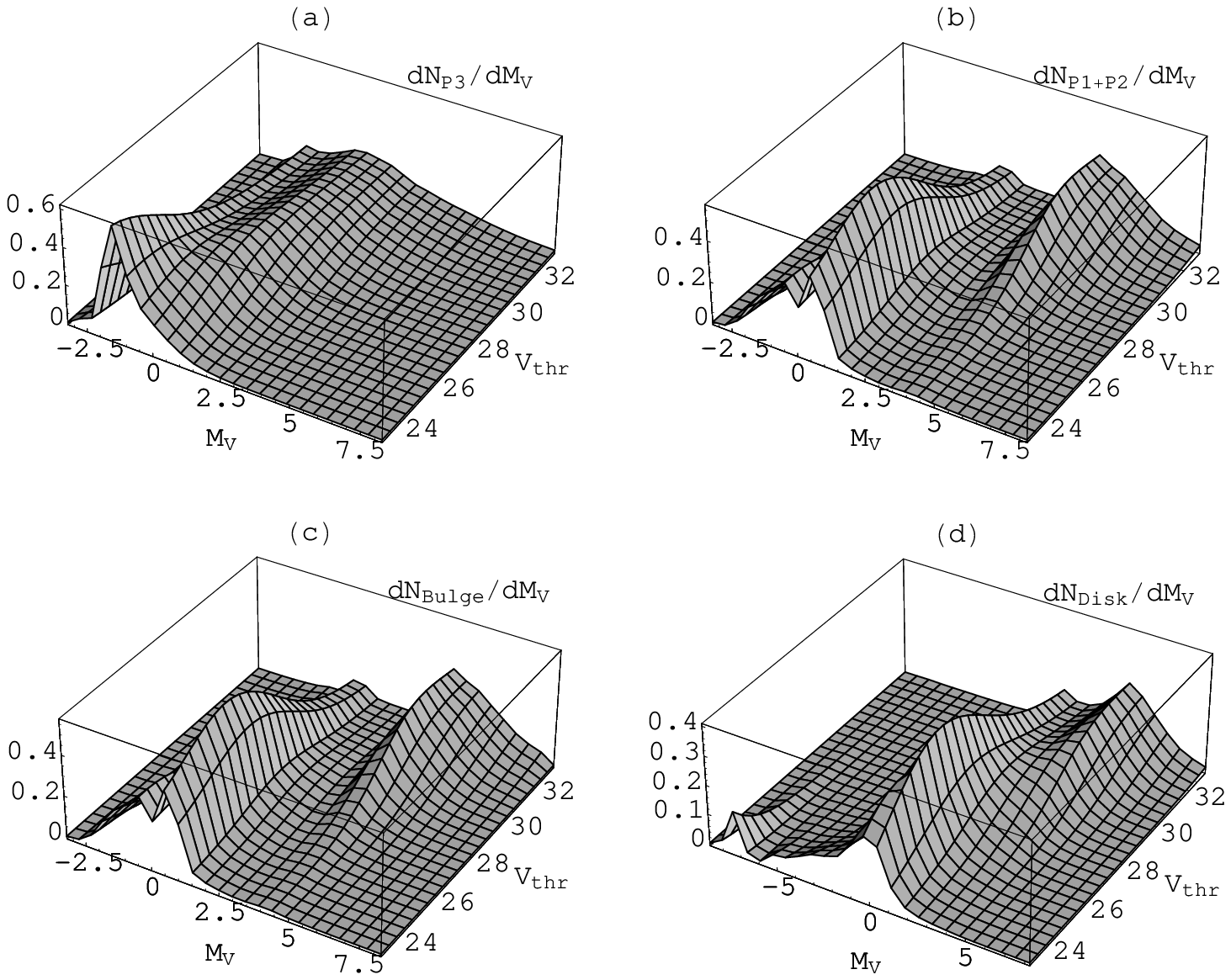}}
 \caption{The same as Figure \ref{Fig dNdK} in the $V$-band.
}
 \label{Fig dNdV}
 \end{center}
\end{figure*}

The number of expected events as calculated in the previous
section is obtained by integrating over the whole range of
possible magnitudes for the sources. It is interesting to evaluate
the contributions of sources with different magnitudes to the
final result. In Figure \ref{Fig dNdK} we show the integrand of
equation (\ref{NPj}) for all four populations as a function of the
source magnitude $M_K$ and the threshold magnitude
$K_{\mathrm{thr}}$. For each value of $K_\mathrm{thr}$ the
function $dN_{P_j}/dM_K$ has been normalized to unity. Figure
\ref{Fig dNdV} shows the same distributions in the $V$-band.

Schematically, we can say that low luminosity stars are more
numerous, whereas brighter stars enjoy a larger lensing zone and
consequently a larger lensing probability. The surfaces in Figures
\ref{Fig dNdK} and \ref{Fig dNdV} are the outcome of the interplay
of these two opposite tensions. At lower thresholds, the
distributions are peaked on the most luminous stars. As we
increase the threshold, less luminous populations become
predominant.

As regards the $K$-band, in the distributions shown in Figure
\ref{Fig dNdK} we can clearly identify the tip of the RGB in the
high luminosity peak dominating at lower values of the threshold.
At intermediate values of $K_\mathrm{thr}$, the red clump gives
the highest contribution. At fainter thresholds, the stars at the
TOP become significant and finally overtake the red clump. It is
interesting to note that the TOP peak gradually shifts to dimmer
values of $M_K$ as $K_\mathrm{thr}$ is increased and less luminous
stars of the main sequence come into play. In the disk
distribution, the brighter main sequence stars and the RSG also
contribute to the first peak at low threshold values.

In the $V$-band, the difference between the four populations is
made more manifest. We can note that P1+P2 and the bulge are
characterized by just two peaks: the one at higher luminosities is
due to the red clump at $M_V\simeq 0$ and the dimmer at $M_V\simeq
4$ is made up of stars at the TOP. The RGB is almost horizontal
and therefore there is no other peak at lower values of $M_V$. In
P3 the red clump and the TOP give rise to two consecutive peaks at
$M_V\simeq -1.5$ and $M_V\simeq 0$ respectively. As the threshold
is increased, the distribution flattens because of the
contribution of progressively fainter stars coming into play. The
disk distributions start with the peak at $M_V\simeq -8$ due to
the RSG, which are absent in the other populations. These stars
give rise to the different behavior of the disk at low thresholds,
already discussed in the previous section. At intermediate
thresholds the red clump becomes dominant and is later followed by
the TOP contribution, which first forms a shoulder to the red
clump peak and then becomes dominant.

Another important point can be anticipated noting that the
distributions for P1+P2 and the bulge are always practically
identical. This proves that differences in the spatial
distribution have very little influence in selecting the class of
sources (luminous or dim) for gravitational lensing. This point
will be better explained in the next section, where we show that
the magnitude and distance distributions can be factorized in a
first approximation.

As a final consideration, we see that the contribution of stars at
$M_K\simeq 6$ or $M_V\simeq 8$ is always negligible for the
thresholds considered in this paper. This provides an {\it a
posteriori} justification of the irrelevance of the faint end
issue of the luminosity function in our calculations, as
anticipated in \S~\ref{Sec Source}.

\section{Distance of the lensed sources}  \label{Sec Dis}

The four source populations considered in this paper fill
different regions of M31, at various distances from the center. It
is interesting to calculate the distribution of the sources
suffering gravitational lensing as a function of their distance
from the central black hole $D_\mathrm{LS}$. Actually, since the
distributions are spread over several orders of magnitudes in
distance, we prefer to present the distributions in
$\log_{10}D_\mathrm{LS}$. In order to get this distribution for
each population, we return to equations (\ref{PiPj}) and
(\ref{NPj}) and reverse the order of integration in
$D_\mathrm{LS}$ and $M_K$
\begin{equation}
\frac{dN_{P_j}}{d\log_{10}D_\mathrm{LS}}\propto D_\mathrm{LS}\int
dM_K ~ n_{P_j}(M_K) \int\limits_0^{R_{\mathrm{Z}}} dr ~ r
\int\limits_0^{2\pi} d\phi f_{P_j}(D_\mathrm{LS},r,\phi).
\label{dNdDLS}
\end{equation}

\begin{figure}[t]
\begin{center}
\resizebox{10cm}{!}{\includegraphics{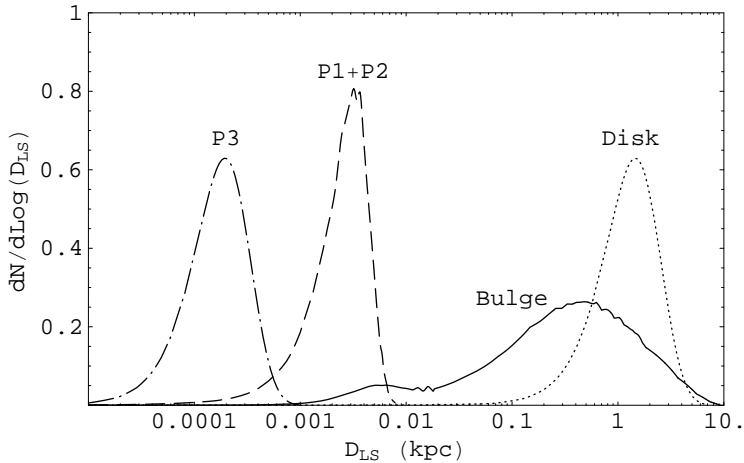}}
 \caption{Distributions for the distance of the lensed sources from the black hole.
}
 \label{Fig dNdDLS}
 \end{center}
\end{figure}

The distributions so-obtained are shown in Figure \ref{Fig
dNdDLS}, after having been normalized to unity. The plots are
obtained for $K_\mathrm{thr}=24$, but the distributions for
different thresholds and even for different bands are practically
indistinguishable. This is a consequence of the fact that the
radius of the lensing zone $R_\mathrm{Z}$ is always much smaller
than the scale of variation of all spatial distributions. Then it
is possible to approximate the spatial distributions in the
integrals (\ref{PiPj}) and (\ref{dNdDLS}) with the values they
assume at the center of the disk of radius $R_\mathrm{Z}$ and
replace the integral over $r$ and $\phi$ by $\pi R_\mathrm{Z}^2
f_{P_j}(D_\mathrm{LS},0,0)$. Since the dependence on
$D_\mathrm{LS}$ in the radius of the lensing zone factors out,
being always $\sqrt{D_\mathrm{LS}}$, the spatial distribution and
the magnitude distributions are effectively factorized. This
decoupling is not possible when we consider the angular position
and the velocity of the images, as will be shown in the succeeding
sections.

The maximum lensing probability is reached by P3 at
$D_\mathrm{LS}=0.19$ pc, by P1+P2 at $D_\mathrm{LS}=3.7$ pc, by
the bulge at $D_\mathrm{LS}=0.48$ kpc and by the disk at
$D_\mathrm{LS}=1.5$ kpc. With respect to the other populations,
the distribution of the bulge events is less localized and
presents a first subpeak at 5 pc. This is a consequence of the
fact that although the bulge is sharply peaked at the center of
the galaxy, it has a long tail decreasing at a lower rate with
respect to the exponential tail characterizing the other
distributions. Moreover, the other populations are more or less
flattened on planes that never contain the line of sight, whereas
the bulge has a more spheroidal shape.

\section{Position of the images}  \label{Sec Ima}

From the observational point of view, it is very important to
determine the expected angular distance from the central black
hole of the secondary images we are looking for. In practice, the
choice of the facilities to employ in the search for these images
is heavily influenced by the resolution required to resolve them.
In this section, we derive the distribution of the secondary
images generated by gravitational lensing with respect to their
angular distance from the central black hole.

Suppose we want to calculate the fraction of sources belonging to
a given population that generate a secondary image at an angular
distance in the range $[\theta,\theta+d \theta]$. The angular
position of these sources is determined through equation
(\ref{LensEq}) as a function of $\theta$ and $D_\mathrm{LS}$. We
thus have to sum up the contributions of sources at different
$D_\mathrm{LS}$. Considering that $|\beta(\theta,D_\mathrm{LS})|$
is a growing function of $D_\mathrm{LS}$, the first contribution
comes from a source at distance $D_\mathrm{min}(\theta)$ such that
$|\beta(\theta,D_\mathrm{min})|=0$. This distance can be
explicitly calculated using equations (\ref{LensEq}) and
(\ref{ThetaEin}). The last contribution comes from sources at the
border of the lensing zone, whose distance
$D_\mathrm{max}(\theta)$ is such that
$|\beta(\theta,D_\mathrm{max})|=\beta_\mathrm{Z}(M_K,K_\mathrm{thr},D_\mathrm{max})$.
This distance is also easily calculable (note that the absolute
value is necessary because we are considering secondary images,
i.e. images with $\theta<\theta_\mathrm{E}$). Finally, for each
source distance in the range $[D_\mathrm{min},D_\mathrm{max}]$, we
must sum up the contributions of all space elements lying on a
circle of radius $D_\mathrm{OS}|\beta(\theta,D_\mathrm{LS})|$
centered on the optical axis at distance $D_\mathrm{LS}$ from the
black hole.

As in the previous section, we prefer to calculate the
distribution in $\log_{10} \theta$. Taking proper account of all
the Jacobians, we have
\begin{equation}
\frac{dN_{P_j}}{d\log_{10}\theta}\propto \theta \int dM_K ~
n_{P_j}(M_K) \int\limits_{D_\mathrm{min}}^{D_\mathrm{max}}
dD_\mathrm{LS} ~ D_\mathrm{OS}^2
|\beta(\theta,D_\mathrm{LS})|\left(2+\frac{|\beta(\theta,D_\mathrm{LS})|}{\theta}
\right) \int\limits_0^{2\pi} d\phi
f_{P_j}(D_\mathrm{LS},\beta,\phi). \label{dNdtheta}
\end{equation}

Equation (\ref{dNdtheta}) is for observations in the $K$-band. The
distribution for observations in the $V$-band can be calculated
similarly with the obvious replacements. Note that now it is
impossible to factor the dependence on $D_\mathrm{LS}$,
$K_\mathrm{thr}$ and $M_K$ as done in the previous section. As a
consequence, the distribution of the angular positions of the
images strongly depends on the threshold magnitude chosen.

\begin{figure*}
\begin{center}
\resizebox{\hsize}{!}{\includegraphics{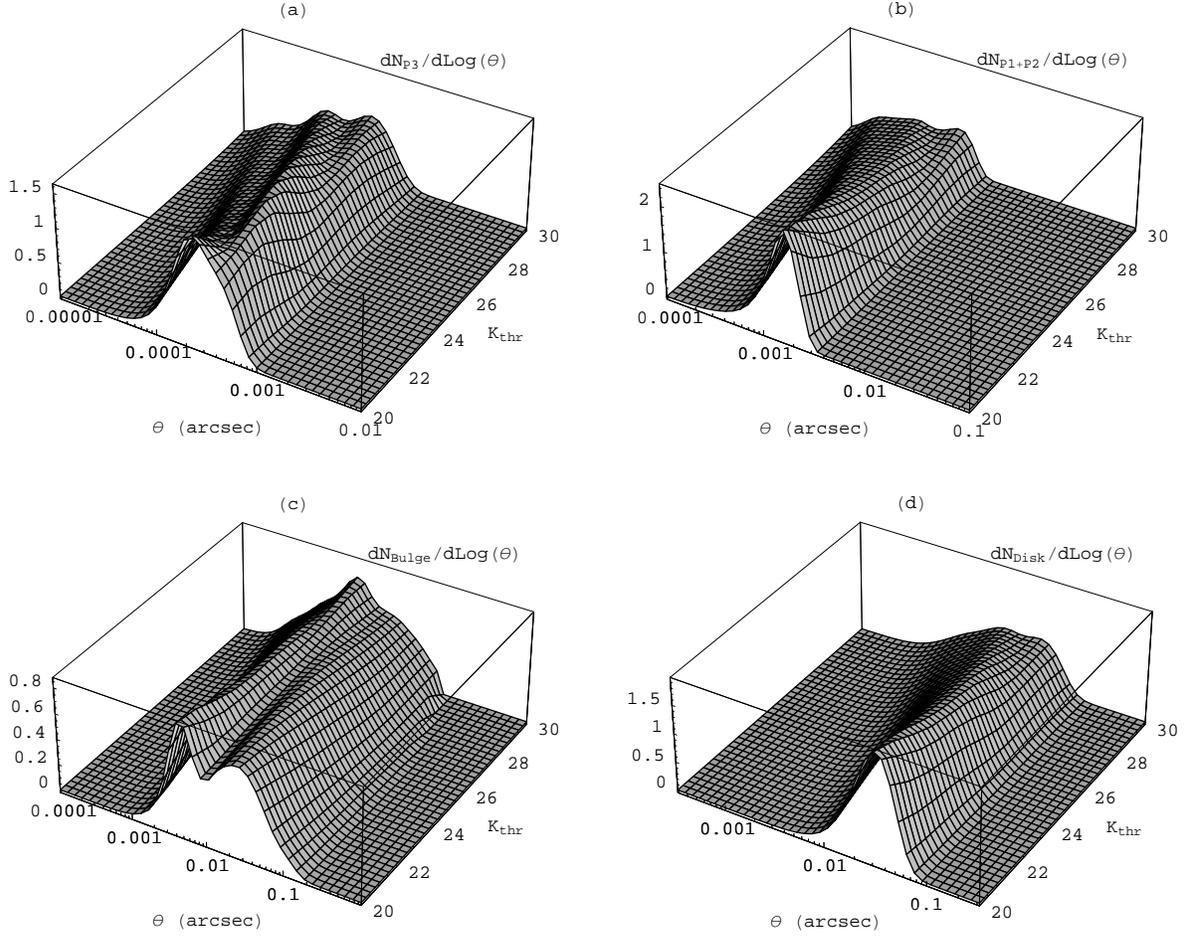}}
 \caption{Distributions of the angular distances of the gravitational lensing
 secondary images from the black hole for the four source populations at different values of the
 threshold magnitude in the $K$-band. (a) P3; (b) P1+P2; (c) bulge; (d) disk. }
 \label{Fig dNdthetaK}
 \end{center}
\end{figure*}

\begin{figure*}
\begin{center}
\resizebox{\hsize}{!}{\includegraphics{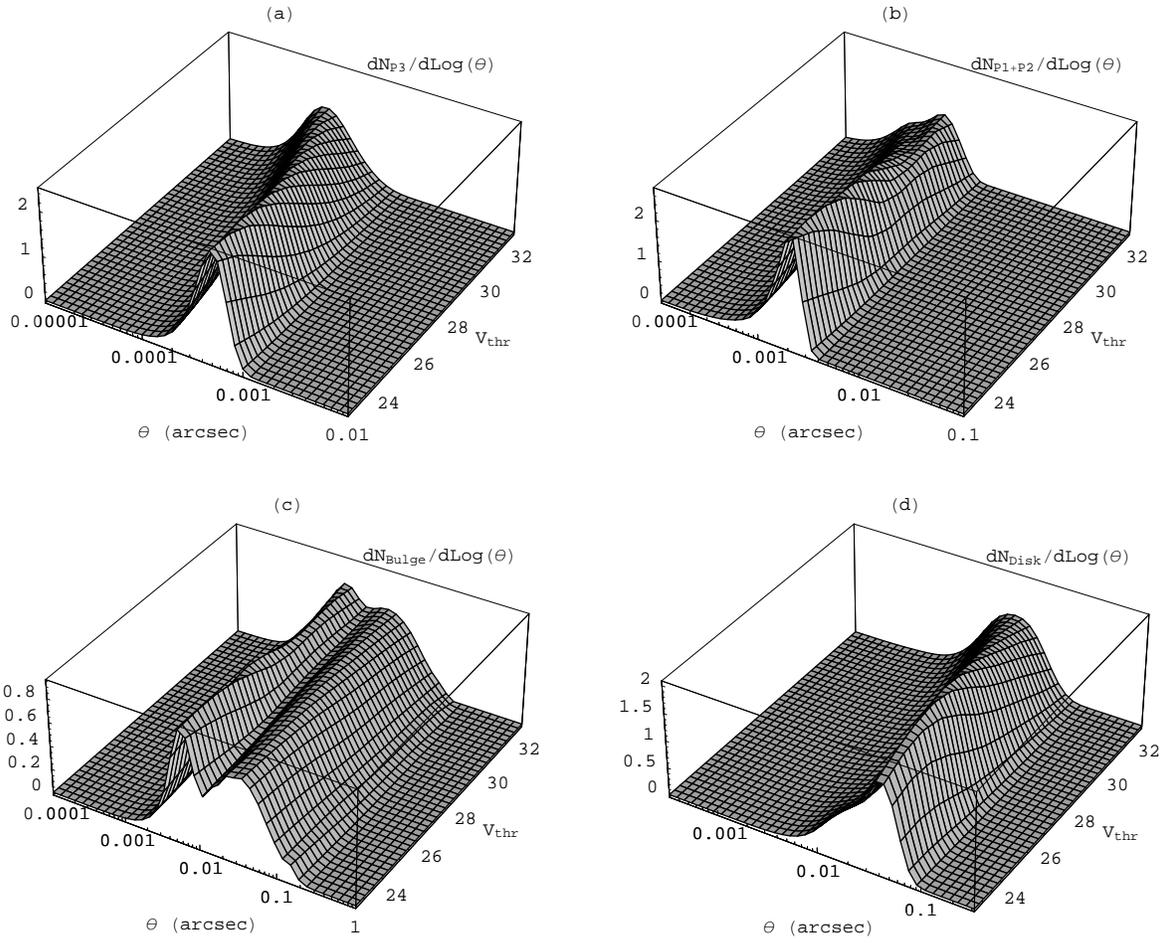}}
 \caption{The same as Figure \ref{Fig dNdthetaK} in the $V$-band. }
 \label{Fig dNdthetaV}
 \end{center}
\end{figure*}

In Figures \ref{Fig dNdthetaK} and \ref{Fig dNdthetaV}  we show
the distributions (normalized to unity at each threshold) for each
source population and for values of the threshold magnitude
ranging from 20 to 30 in the $K$-band and from 23 to 33 in the
$V$-band respectively. The images of sources belonging to P3 have
angular distance in the range $[0.0001'',0.001'']$, those
belonging to P1+P2 appear at a few mas from the central black
hole, those in the disk have angular distance in the range
$[0.01'',0.1'']$, whereas sources in the bulge can generate images
with a larger spread of angular distances, from $0.001''$ to
$0.1''$. In general, as we increase the threshold magnitude, the
distributions move to slightly lower values of $\theta$. This can
be understood by the fact that lowering the threshold we accept
fainter images generated by sources more distant from the optical
axis, whose secondary images appear closer to the black hole. In
practice, as a general rule, we learn that, in order to catch all
fainter images one gets by performing deeper observations, one
also needs better resolution. This shift is partly compensated by
the existence of numerous intrinsically dim stars, which anyway
need a very good alignment in order to generate a visible
secondary image. These dim stars tend to keep the distribution at
the highest possible value of $\theta$, essentially fixed by the
Einstein angle at the typical distance scale of the population.

Apart from this general behavior, we can clearly identify some of
the populations discussed in \S~\ref{Sec Mag}. For example, in the
distribution for P3 in the $K$-band we can clearly distinguish
three peaks at high threshold, corresponding to the tip of the
RGB, the red clump and the TOP from left to right. These peaks are
more smoothed in the distributions for P1+P2 and the disk. In the
bulge, the double peak structure is mainly determined by its
spatial distribution, which presents a long tail extending up to a
few kpc. This long tail is responsible for the peak at $\theta
\simeq 0.01''$. This can be seen by the fact that at
$K_\mathrm{thr}=20$, at which we expect a single peak due to the
tip of the RGB, we already have two evident peaks. Moreover, the
same structure can be seen in the $V$-band distribution,
notwithstanding the different magnitude distribution.

The $V$-band distributions appear to be less structured with just
a very slight shift to lower values of $\theta$ with the increase
of $V_\mathrm{thr}$.

Summing up, we note that imaging the central regions of the
nucleus of M31 with a resolution of the order of a few mas, as
could be possible with the Keck or the LBT, will allow to catch
secondary images of sources in the disk and a large part of those
in the bulge, provided one reaches a faint enough $K_\mathrm{thr}$
(see discussion in \S~\ref{Sec ID}).

\section{Velocity of the images} \label{Sec Vel}

Up to now, we have just calculated the number of gravitational
lensing events that are observable at a given time. It is very
important to understand the timescale of the variations of the
geometric configuration of such lensing events. In fact, the
strategies for the observations tightly depend on these timescales
and can be very different for static or nearly static
configurations as opposed to events with any secular development.
As will be discussed in \S~\ref{Sec ID}, the detection of the
motion of the images might be of key importance to identify
genuine gravitational lensing events. This section is devoted to
the presentation of the distributions of the expected velocities
of the secondary images below a given threshold magnitude.

These distributions are obtained by a Montecarlo procedure. For
each value of $M_K$ and $K_\mathrm{thr}$, we have generated $10^4$
stars, randomly choosing their position within the lensing zone
corresponding to these magnitudes. The velocities of these sources
have been randomly extracted from a two-dimensional gaussian
distribution centered on the rotation velocity $v_{P_j}$ with
dispersion $\sigma_{P_j}$. The values of these parameters are
specified in \S~\ref{Sec Source} for each population.

Once the sources have been generated, we have calculated the
apparent proper motions of the corresponding secondary images by
differentiating the standard formula for the position of the
secondary image. The result is

\begin{equation}
v_\theta = \left| \frac{d
\vec{\theta}_-}{dt}\right|=\frac{\sqrt{u^2+4\sin^2 \phi}}{2}
\left(\frac{1}{u}-\frac{1}{\sqrt{u^2+4}} \right)
\frac{v}{D_\mathrm{OS}}, \label{vtheta}
\end{equation}
where $v_\theta$ is the modulus of the time derivative of the
angular position of the secondary image, $v$ is the modulus of the
transverse velocity of the source and $\phi$ is the angle between
the velocity and position vectors of the source.

The image velocity decreases as $u^{-2}$ when the source is far
away from the optical axis. In this limit, the secondary image is
also very faint and close to the black hole. In the opposite
limit, when $u\rightarrow 0$, the velocity diverges (except for
the case $\phi=0$). This is very well-known in microlensing
studies, as the secondary image moves very rapidly when the source
is at the closest approach distance.

\begin{figure*}
\begin{center}
\resizebox{\hsize}{!}{\includegraphics{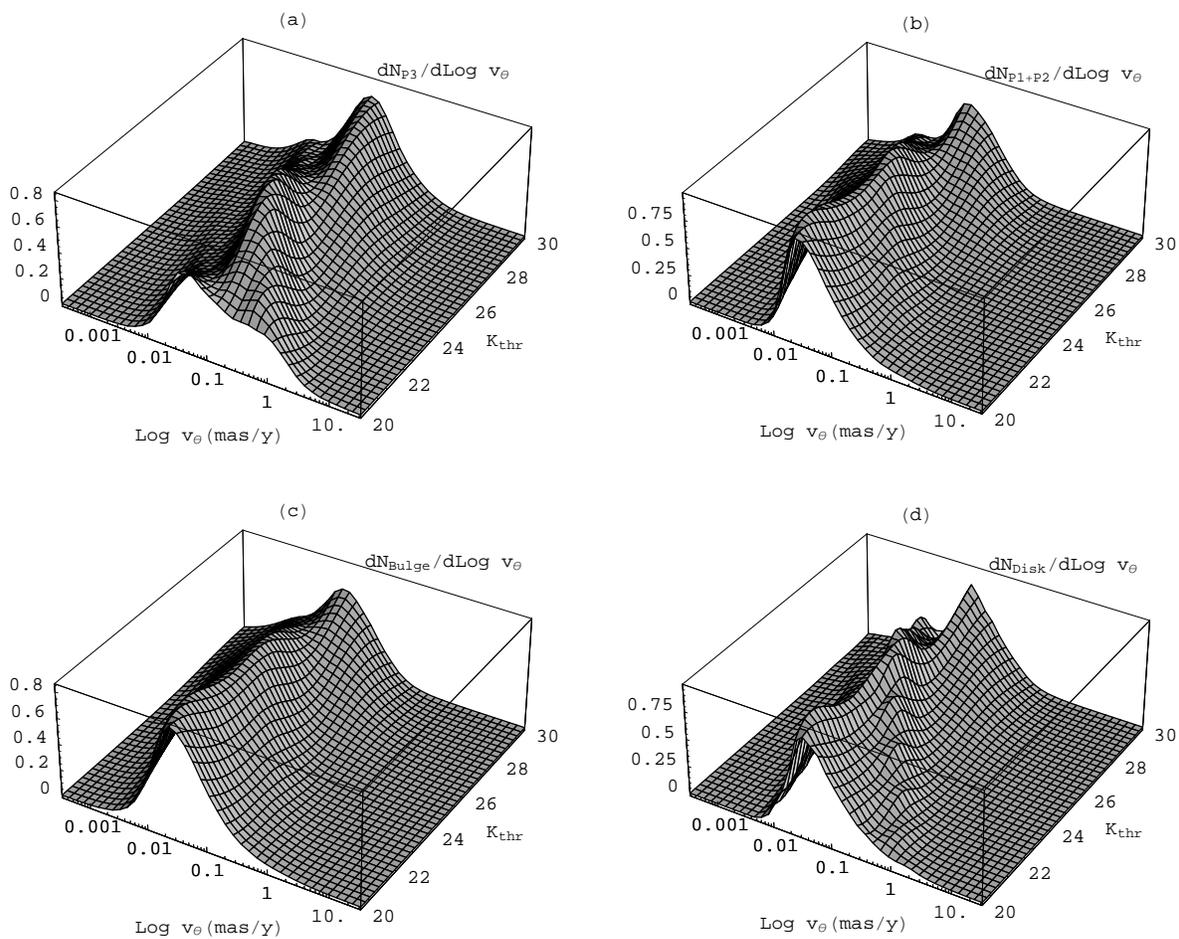}}
 \caption{Distributions of the velocities of the gravitational lensing
 secondary images from the black hole for the four source populations at different values of the
 threshold magnitude in the $K$-band. (a) P3; (b) P1+P2; (c) bulge; (d) disk. }
 \label{Fig dNdvK}
 \end{center}
\end{figure*}

\begin{figure*}
\begin{center}
\resizebox{\hsize}{!}{\includegraphics{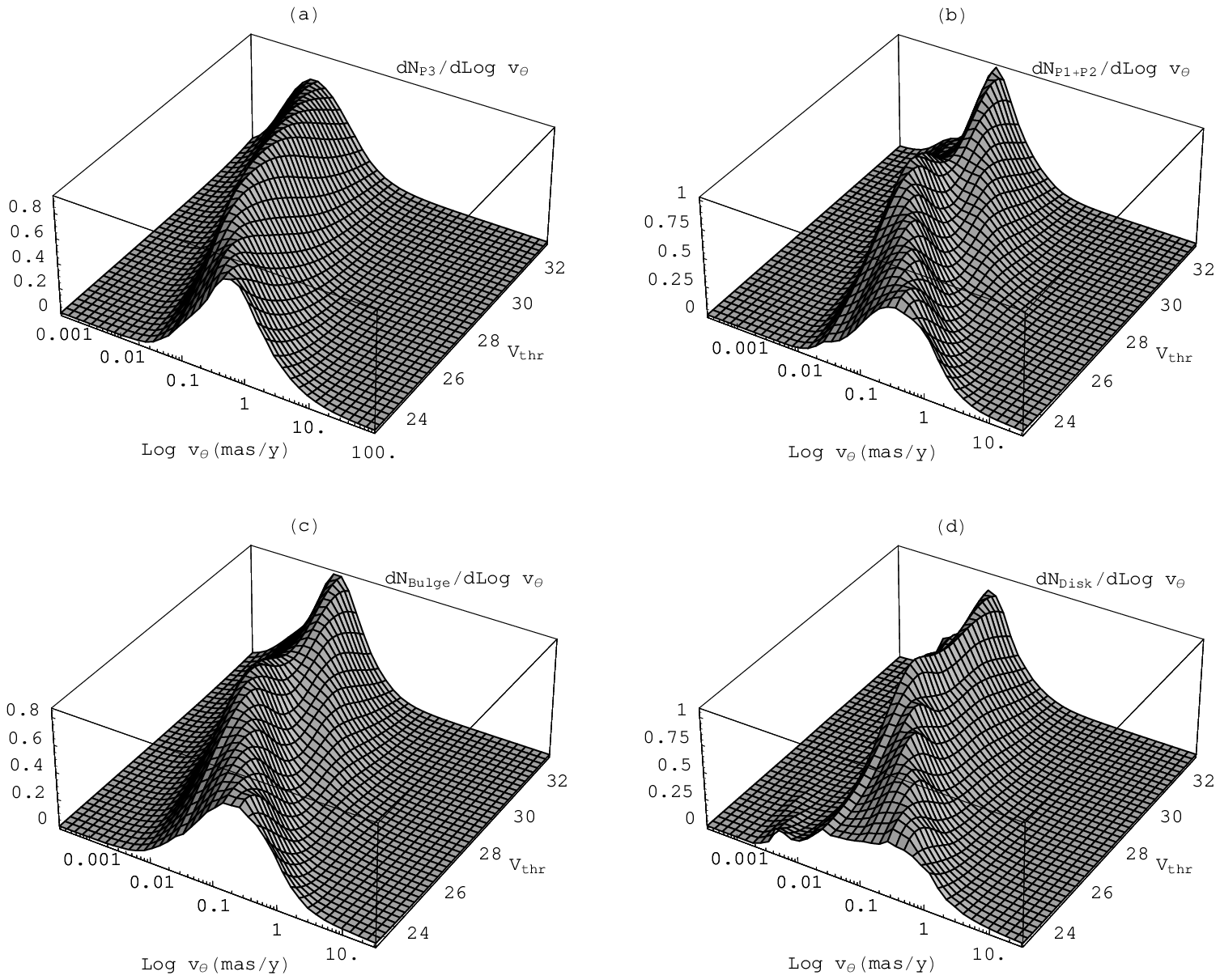}}
 \caption{The same as Figure \ref{Fig dNdvK} in the $V$-band. }
 \label{Fig dNdvV}
 \end{center}
\end{figure*}

With the velocities of the images thus calculated, we have
constructed the distributions shown in Figures \ref{Fig dNdvK} and
\ref{Fig dNdvV} for the $K$-band and $V$-band respectively. The
proper motion has been expressed in milliarcseconds per year.

The expected apparent proper motions of the secondary images cover
a very large range. All four populations generate images whose
velocities generally lie between $0.01$ and $0.1$ mas$/$y, with
very long tails extending one order of magnitude above and below
these limits.

Going from lower to higher thresholds, all distributions
significantly drift to lower values of the proper motion. This is
clear from the fact that, at bright thresholds, very good
alignments generate higher velocities as it is evident from
equation (\ref{vtheta}) and viceversa. Then, going to fainter
thresholds, the lensing zone is enlarged, and less aligned sources
participate in the distribution pulling it to lower values of
$v_\theta$. This is particularly evident in the $V$-band.

Thanks to the stronger dependence of the velocity distributions on
the threshold magnitude, the internal structure of the four
populations in terms of stellar components is made more evident.
In the $K$-band we can clearly identify three peaks moving from
right to left as $K_\mathrm{thr}$ is increased from 20 to 30. As
usual, the peak on the left is due to the tip of the RGB, the
central peak is due to the red clump and the one on the right is
made up of TOP stars, coming into play at higher thresholds. In
the $V$-band there is a first peak formed by red clump stars
moving from high to low values of $v_\theta$. At high thresholds
the TOP stars are able to form a shoulder and then a second peak
on the right of the red clump peak. At low values of
$V_\mathrm{thr}$ the disk distribution is influenced by RSG, which
form the tail at small $v_\theta$. The contribution of RSG becomes
subdominant already at intermediate thresholds, where it is taken
over by red clump stars and finally by TOP stars.

\section{Rate and duration of the events} \label{Sec Kin}

An interesting quantity for the definition of the timescale of
observational campaigns is the rate of the events, defined as the
number of new detectable events occurring per unit time. It can be
calculated by
\begin{equation}
\Gamma_{P_j}=\int dM_K ~ n_{P_j} \int\limits_0^\infty
dD_\mathrm{LS} ~ 2\int\limits_0^{R_\mathrm{Z}} dr
\frac{1}{2\pi}\int\limits_0^{2\pi} d\phi
 ~ f_{P_j} ~ \sigma_{P_j} ~ G\left( v_{P_j}/\sigma_{P_j}\right),
 \label{Rate}
\end{equation}
where
\begin{equation}
G(x)=\sqrt{\frac{\pi}{8}}e^{-x^2/4}\left[(2+x^2)I_0(x^2/4)+x^2I_1(x^2/4)
\right]
\end{equation}
and $I_\nu$ is the modified Bessel function of order $\nu$.

The combination $\sigma_{P_j} ~ G\left(
v_{P_j}/\sigma_{P_j}\right)$ is the result of the integration on
the velocity distribution, again assumed to be a gaussian with
dispersion $\sigma_{P_j}$ centered on the rotation velocity
$v_{P_j}$. For further details on this point, see the appendix of
\citet{AleSte}.

The function $G(x)$ satisfies the following limits
\begin{eqnarray}
&& \lim\limits_{v_{P_j}\rightarrow 0}\sigma_{P_j} ~ G\left(
v_{P_j}/\sigma_{P_j}\right) =\sqrt{\pi/2}~\sigma_{P_j}\\
&& \lim\limits_{\sigma_{P_j}\rightarrow 0}\sigma_{P_j} ~ G\left(
v_{P_j}/\sigma_{P_j}\right) =v_{P_j}.
\end{eqnarray}

\begin{figure}
\begin{center}
\resizebox{10cm}{!}{\includegraphics{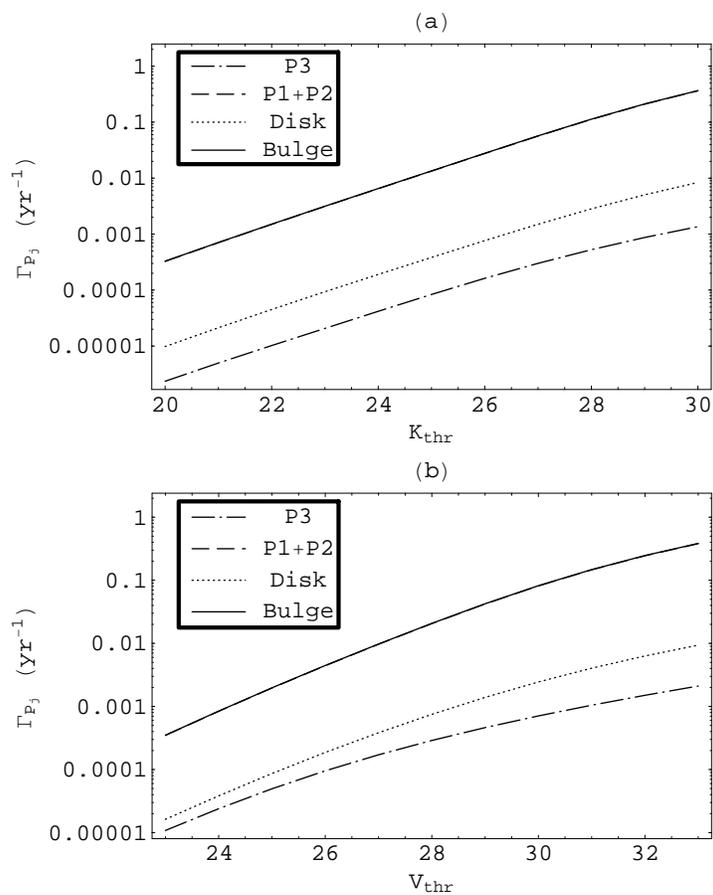}}
 \caption{(a) Rate of the gravitational lensing events for the four
 source populations considered in the text expected in the $K$-band as function
 of the threshold magnitude. (b) The same in the $V$-band. The
 rate of P1+P2 and the rate of the bulge practically coincide.
}
 \label{Fig Rate}
 \end{center}
\end{figure}

Applying equation (\ref{Rate}) to each of our four populations, we
obtain the curves in Figure \ref{Fig Rate}. We see that the more
distant populations move more slowly and thus generate new events
at a lower rate. For this reason the rate of the bulge events
practically coincides with the rate of P1+P2, while the rate of
the disk drops to levels comparable to those of P3.

Having calculated both the rate and the number of simultaneously
present events, we immediately derive the average time spent by an
event above threshold as
\begin{equation}
<\Delta T_{P_j}>= \frac{N_{P_j}}{\Gamma_{P_j}}\simeq \frac{\pi}{2}
\frac{<R_\mathrm{Z}>}{<v_{\perp}>}, \label{DeltaT}
\end{equation}
where the last equality holds approximatively and involves the
average radius of the lensing zone $<R_\mathrm{Z}>$, and the
average transverse velocity $<v_{\perp}>$.

\begin{figure}
\begin{center}
\resizebox{10cm}{!}{\includegraphics{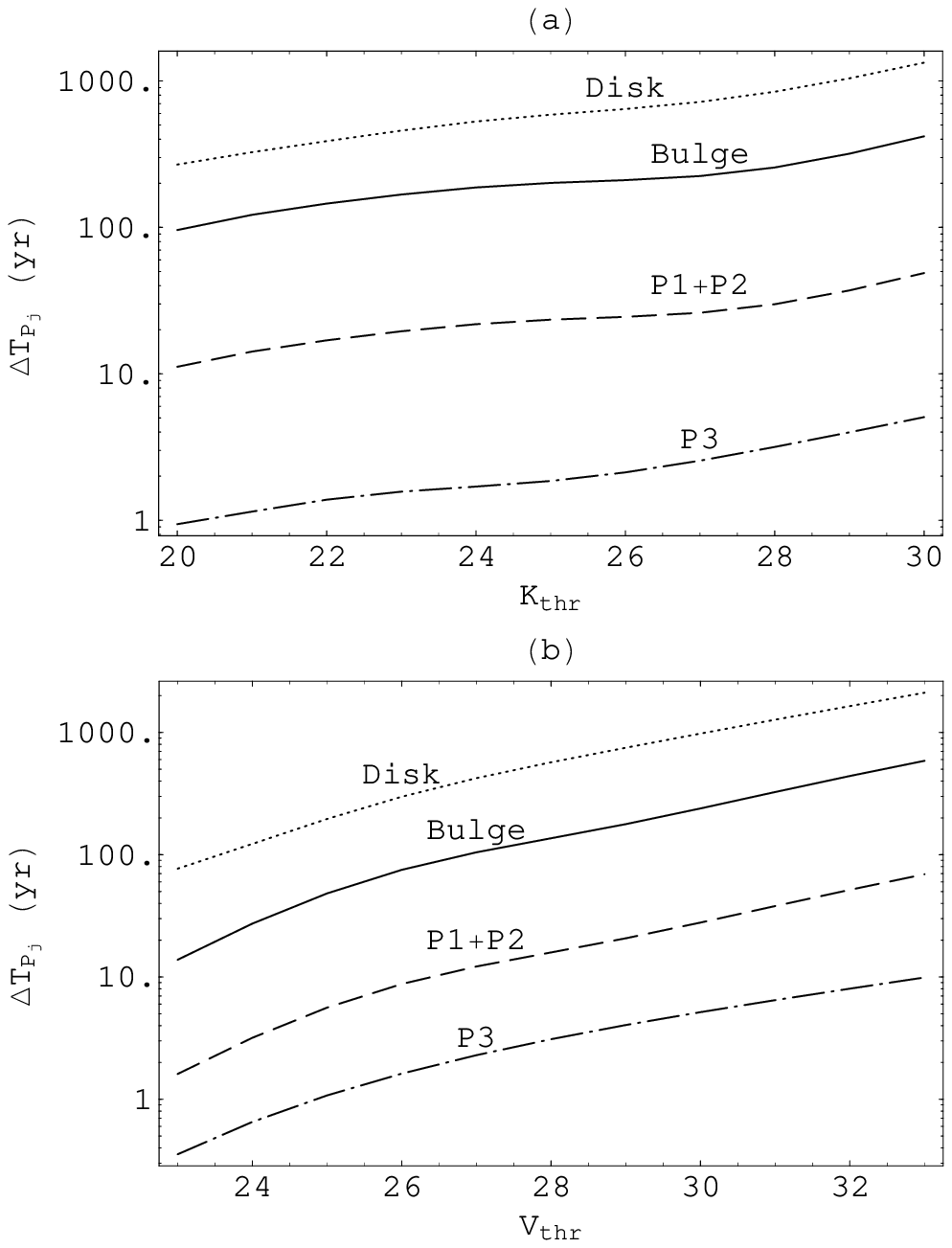}}
 \caption{(a) Average time spent above threshold in the $K$-band for the gravitational lensing events for the four
 source populations. (b) The same in the $V$-band.
}
 \label{Fig DeltaT}
 \end{center}
\end{figure}

The average duration of the events for each source population are
drawn in Figure \ref{Fig DeltaT}. It grows with the chosen
threshold of the experiment. This is due to the fact that as
$K_\mathrm{thr}$ is increased the lensing zone becomes larger and
larger and the time required by the sources to cross it becomes
proportionally larger (see the approximate expression in eq.
[\ref{DeltaT}]).

The modulation in the growth at intermediate thresholds in the
$K$-band is again an effect of the presence of different stellar
branches in each population. In fact, at low $K_\mathrm{thr}$ most
of the events are due to bright sources in the red clump. Given
their intrinsic luminosity, these sources generate secondary
images that stay longer above threshold. At intermediate
thresholds, there emerges a non-negligible contribution from
fainter sources. These sources have a smaller lensing zone and
their secondary images spend much less time above threshold with
respect to the brighter sources. The distribution of the average
duration becomes bimodal, with the brighter sources enjoying a
long time above threshold and the fainter sources spending less
time above threshold. The average time of such a bimodal
distribution is therefore in the middle between the two maxima. As
we increase $K_\mathrm{thr}$ further, the fainter sources dominate
with respect to the brighter and $\Delta T_{P_j}$ is determined by
these sources only. Then the modulation observed at intermediate
$K_\mathrm{thr}$ in the curves in Figure \ref{Fig DeltaT} is just
the transition from a regime dominated by more luminous sources to
a regime dominated by dimmer sources. In the $V$-band this
transition is less evident.

Coming to a quantitative analysis of the average duration, we note
that the events involving sources in the disk of M31 last up to
several centuries. For the bulge the average time spent above
threshold ranges from 50 to 200 years. For P1+P2 the situation is
more dynamical, thanks to the higher velocities characterizing the
stars very close to the central black hole. An observational
campaign lasting several years could study the evolution of the
secondary images already present and hope to see new events (if
the threshold magnitude is sufficiently high). The average
duration of the events involving sources in P3 is of the order of
one year, but the extremely low rate of new events for this
population leads one to discard it as a source of gravitational
lensing events.

\section{Identification of the lensing images} \label{Sec ID}

On the basis of the analysis given in the previous sections, it is
time to set up a realistic methodology to select the candidate
lensing images and test their authenticity.

As mentioned before, the most probable lensing events will involve
bulge stars as sources, at typical distances ranging from a few pc
to 1 kpc from the lens. In this configuration, the secondary
images may form at an angular distance from the central black hole
ranging from a few mas to $0.1''$. In order to get access to the
bulk of the gravitational lensing events, it is then mandatory to
employ interferometry. At this point, the question arises whether
a secondary image of a lensed star can be distinguished from a
background star unambiguously.

In this respect, it is instructive to calculate the expected
number of background stars in an angular area within radius
$\theta$ centered on the black hole. For a threshold magnitude
$K_{\mathrm{thr}}$, only the sources with observed magnitude
$K<K_\mathrm{thr}$ will be observable. This fixes a threshold on
the intrinsic magnitude of the sources as
\begin{equation}
M_{K,\mathrm{max}}=K_\mathrm{thr}-5\log_{10}\frac{D_{\mathrm{OS}}}{10\mathrm{~
pc}}-A_K.
\end{equation}

The number of background sources is then
\begin{equation}
B_{P_j}(K_\mathrm{thr},\theta)=\int\limits_{-\infty}^{M_{K,\mathrm{max}}}
dM_K \; n_{P_j}(M_K) \int\limits_{-D_\mathrm{OL}}^\infty d z
\int\limits_0^{\theta (z+D_\mathrm{OL})} dr ~ r
\int\limits_0^{2\pi} d\phi f_{P_j}(z,r,\phi). \label{BPj}
\end{equation}

Actually, since the spatial distributions for all source
populations are centered on the black hole and have an extent much
smaller than $D_\mathrm{OL}$, we can take the approximation $z\ll
D_{\mathrm{OL}}$ and push the lower extremum in the $z$-integral
to $-\infty$.

Let us imagine an observational campaign to discover secondary
images due to gravitational lensing lying within $\theta=0.01''$
of the central black hole. Fixing $K_\mathrm{thr}=24$, the number
of background stars in such area is $B_\mathrm{P3}=3.5\times
10^{-2}$, $B_\mathrm{P1+P2}=8.3$, $B_\mathrm{Bulge}=3.7$,
$B_\mathrm{Disk}=1.1\times 10^{-2}$.

Our image will thus contain something like 12 stars within
$0.01''$ from the central black hole. The first step would be to
check for possible alignments between these stars and the central
black hole. If we find that the line joining two stars passes
through the central black hole, we can select these stars as
candidate lensing images of the same source. Since gravitational
lensing is an achromatic phenomenon, the two images must have the
same color indices (unless there is reason to believe that the
photon paths of the two images cross regions with sensibly
different amount of dust). Therefore, false candidates could be
excluded taking images in different IR bands and checking whether
the luminosity ratio between the two images remains the same in
all bands.

A further test is to take another image after several months or
even the next year. If the alignment is just a chance product and
one of the two stars has a high enough proper motion (which is
likely to happen if the star is intrinsically close to the central
black hole), then the alignment will be lost in the second image.
Conversely, if the alignment is an authentic product of
gravitational lensing, and if any evolution shows up, then the two
images must move in such a way that the line joining them always
passes through the black hole. However, as shown in in \S~\ref{Sec
Vel}, the proper motion of the secondary images is typically below
$0.1$ mas$/$y. Therefore, it would be improbable that the observed
events show any evolution within less than ten years. Moreover,
given the very low rate of new events, follow-up observations
would make sense only if thresholds higher than
$K_\mathrm{thr}=30$ are reached in the observations.

Our estimate of 12 background stars in an area of radius $0.01''$
centered on the black hole is based on our spatial distributions,
which do not take into account a possible presence of a further
cuspy cluster of stars internal to P3 and thus very close to the
supermassive black hole. If such a cluster were present, then the
number of background stars would sensibly increase. However, these
stars would move very fast and chance alignments in a first image
would be easily discarded after the analysis of a second image. As
stressed before, the most delicate background actually comes from
stars that lie along the line of sight but intrinsically far from
the black hole and consequently have slow proper motion. We
believe that the number of these stars is correctly estimated
through equation (\ref{BPj}). For completeness, we also mention
that the estimate of background stars for $K_\mathrm{thr}=30$
yields 1600 stars and for $V_\mathrm{thr}=33$ gives 1320 stars.

As a final consideration, we note that the deformation of the
point-like image of the source is of the order of $10^{-7}$ arcsec
for the gravitational lensing events considered in this work. This
prevents from using image deformation as a selection tool for
lensing effects.

\section{Conclusions}

In this paper we have undertaken a deep investigation of the
possible observation of gravitational lensing effects due to the
black hole in the center of M31. We have considered stars
belonging to four different populations (bulge, disk and the
central clusters P1+P2 and P3) as candidate sources. Through
detailed modelling of the spatial distributions and the luminosity
functions of these four populations we have calculated several
quantities that can be used to quantify the power of the central
black hole as a gravitational lens using present and future
observational facilities. We have carried out our analysis both in
the $K$-band and in the $V$-band.

The main outcome is the number of expected lensed sources at a
given time whose secondary image is brighter than a specified
threshold magnitude. The results are summarized in Figure \ref{Fig
NEvents} and Tables \ref{Tab NEventsK} and \ref{Tab NEventsV},
showing that indeed we expect 1.4 lensed sources at a threshold
$K_\mathrm{thr}=24$, 16 events at $K_\mathrm{thr}=27$ and 180
events at $K_\mathrm{thr}=30$. In the $V$-band we would have 1.3
events at $V_\mathrm{thr}=27$, 25 events at $V_\mathrm{thr}=30$
and 270 events at $V_\mathrm{thr}=33$.

We have also presented the distribution of the lensed sources as a
function of their absolute magnitude, showing the contribution of
different stellar branches as sources of gravitational lensing
events.

The distribution of the lensed sources as a function of the
distance shows that the disk stars are mostly lensed at about 1
kpc distance from the black hole, whereas the bulge stars can be
lensed at distances ranging from a few pc to 1 kpc. The inner
cluster stars are lensed at fractions of a pc.

This difference is reflected in the angular separation of the
images from the black hole, which ranges from 1 mas to $0.1''$ for
bulge and disk stars, whereas it stays of the order 1 mas or below
for the inner clusters. The resolution needed to get a significant
number of events is thus of the order of a few mas, requiring the
employment of long baseline interferometers or extremely large
telescopes. This justifies our choice to present our analysis in
the $K$-band, for which the present infrared interferometers such
as Keck, LBT and the future space telescope JWST are optimized.
The VLTI, though representing the most advanced interferometer
operating in IR bands, cannot observe M31 efficiently. The
parallel analysis in the $V$-band provides a deeper comprehension
of all the effects coming into play. In this band, the
expectations for the number of lensing events are comparable to
those in the $K$-band, provided that $V$-band interferometers are
constructed in the future with sizes of the same order as in the
$K$-band.

The gravitational lensing events discussed in this paper have a
very slow evolution, with typical angular velocities of the images
between $0.01$ and $0.1$ mas$/$y. We have outlined a possible
observational strategy to select gravitational lensing events
within the expected background stars, by checking the alignment of
pairs of images, the achromaticity of the flux ratio and setting
up follow-up observations to detect any secular development.

Our analysis can be repeated with different models of the source
populations. Given the present uncertainties on the morphology of
the components of M31, we believe that our estimates can be
corrected by a factor of a few at most.

The chances for concrete observations of gravitational lensing
events by the supermassive black hole in M31 are rather low
(though not null) even with the best facilities available today.
The perspectives will be definitely increased with the realization
of future projects such as the JWST, the extremely large ground
telescopes or the realization of new long baseline optical
interferometers in the northern hemisphere. With such facilities
it should be possible to reach the higher thresholds indicated in
this paper while keeping a very high angular resolution. This
would open the way to an intensive research of gravitational
lensing events and to their use in the investigation of the
environment of the supermassive black hole.

\begin{acknowledgements}
We are grateful to Gaetano Scarpetta and an anonymous referee for
some fundamental comments which have taken to a consistent
improvement of the manuscript. This work has made use of the
IAC-STAR Synthetic CMD computation code. IAC-STAR is supported and
maintained by the computer division of the Instituto de
Astrof\'isica de Canarias. The authors acknowledge support for
this work by MIUR through PRIN 2006 Prot. 2006023491\_003 and by
research funds of the Salerno University.
\end{acknowledgements}

------------------------------------------------------------------

\bibliographystyle{aa}

\begin{thebibliography}{}

\bibitem[Alexander(2001)]{Ale}  Alexander, T. 2001, \apj, 553,
L149

\bibitem[Alexander \& Loeb(2001)]{AleLoe}  Alexander, T., \& Loeb, A. 2001, \apj, 551,
223

\bibitem[Alexander \& Sternberg(1999)]{AleSte}  Alexander, T., \& Sternberg, A. 1999, \apj, 520,
137

\bibitem[Aparicio \& Gallart(2004)]{ApaGal}  Aparicio, A., \& Gallart, C. 2004, \aj, 128,
1465

\bibitem[Bellazzini et al.(2003)]{Bel03} Bellazzini, M., Cacciari, C., Federici, L., et al. 2003,
\aap, 405, 867

\bibitem[Bender et al.(2005)]{Bender05} Bender, R., Kormendy, J., Bower, G., et al. 2005,
\apj, 631, 280

\bibitem[Bozza(2002)]{Boz1} Bozza, V. 2002, \prd, 66, 103001

\bibitem[Bozza \& Mancini(2004)]{BozMan1}
Bozza, V. \& Mancini, L. 2004, \apj, 611, 1045

\bibitem[Bozza \& Mancini(2005)]{BozMan2}
Bozza, V. \& Mancini, L. 2005, \apj, 627, 790

\bibitem[Chanam\'e et al.(2001)]{CGM}  Chanam\'e, J., Gould, A., \& Miralda-Escud\'e, J. 2001, \apj, 563,
793

\bibitem[Chang et al.(2007)]{CMCQ}  Chang, Ph., Murray-Clay, R., Chiang, E., \& Quataert, E.
2007, \apj, 668, 236

\bibitem[Cox(2001)]{5.77}  Cox, A.N. 2001, Allen's Astrophysical
quantities, New York, Springer

\bibitem[Darwin(1959)]{Dar}  Darwin, C. 1959, Proc. of the Royal Soc. of London A, 249, 180

\bibitem[De Paolis et al.(2003)]{DeP}
De Paolis, F., Geralico, A., Ingrosso, G., \& Nucita, A.A. 2003,
\aap, 409, 809

\bibitem[Demarque \& Virani(2007)]{DemVir}
 Demarque, P., \& Virani, S., 2007, \aap, 461, 651

\bibitem[Eckart \& Genzel(1997)]{EckGen}  Eckart, A., \& Genzel, R. 1997, \mnras, 284,
576
%
\bibitem[Eckart et al.(2002)]{Eckart}  Eckart, A., Genzel, R., Ott, T., \& Sch\"odel, R. 2002, \mnras, 331,
917

\bibitem[Eisenhauer et al.(2005)]{Eis}  Eisenhauer, F., Genzel, R., Alexander, T., et al. 2005,
\apj, 628, 246

\bibitem[Gardner et al.(2006)]{Gar}  Gardner, J.P., Mather, J.C., Clampin, M., et al. 2006,
Space Science Reviews, 123, 485

\bibitem[Genzel et al.(1996)]{Genzel}  Genzel, R., Thatte, N., Krabbe, A., et al. 1996, \apj, 472,
153

\bibitem[Ghez et al.(1998)]{Ghez1}  Ghez, A.M., Klein, B.L., Morris, M., \& Becklin, E.E. 1998, \apj, 509,
678

\bibitem[Ghez et al.(2003)]{Ghez2}  Ghez, A.M., Duch\^ene, G., Matthews, K., et al. 2003, \apj, 586,
L127

\bibitem[Ghez et al.(2005)]{Ghez3} Ghez, A.M., Salim, S., Hornstein,
et al. 2005, \apj, 620, 744

\bibitem[Han(1989)]{Han}  Han, C. 1996, \apj, 472,
108

\bibitem[Hodge(1989)]{Hodge}  Hodge, P. 1989, \araa, 27,
139

\bibitem[Jaroszy\'nski(1998)]{Jar}  Jaroszy\'nski, M., 1998, Acta Astron., 48,
413

\bibitem[Keeton \& Petters(2005)]{KeePet}
 Keeton, C.R., \& Petters, A.O. 2005, \prd, 72, 104006

\bibitem[Kent(1989)]{Kent}  Kent, S.M. 1989, \aj, 97,
1614

\bibitem[King et al.(1995)]{King95} King, I.R., Stanford, S.A., \& Crane, P. 1995,
\aj, 109, 164

\bibitem[Kochanek et al.(2004)]{SaasFee} Kochanek, C.S., Schneider, P., \& Wambsganss, J. 2004, Proceedings of the
33rd Saas-Fee Advanced Course, G. Meylan, P. Jetzer \& P. North,
eds. (Springer-Verlag: Berlin)


\bibitem[Kroupa et al.(1993)]{KTG}  Kroupa, P., Tout, C.A., \& Gilmore, G. 1993, \mnras, 262,
545

\bibitem[Lauer et al.(1993)]{Lauer93} Lauer, T.R., Faber, S.M., Groth, E.J., et al. 1993,
\aj, 106, 1436

\bibitem[Nusser \& Broadhurst(2004)]{NusBro} Nusser, A., \& Broadhurst, T.  2004,
\mnras, 355, L6

\bibitem[Olsen et al.(2006)]{Olsen06} Olsen, K.A.G., Blum, R.D., Stephens, A.W., et al.
2006, \aj, 132, 271

\bibitem[Paumard et al.(2006)]{Paumard} Paumard, T., Genzel R.,
Martins, F., et al. 2006, \apj, 643, 1011

\bibitem[Peiris \& Tremaine(2003)]{PeiTre} Peiris, H.V., \& Tremaine, S. 2003, \apj, 599, 237

\bibitem[Reid et al.(2006)]{Reid} Reid, M.J., Menten, K.M., Trippe, S., et al. 2006,
\apj, 659, 378

\bibitem[Sarajedini \& Jablonka(2005)]{SarJab} Sarajedini, A., \& Jablonka, P. 2005, \aj, 130,
1627

\bibitem[Schlegel et al.(1998)]{Schlegel} Schlegel, D.J., Finkbeiner, D.P., \& Davis, M.  1998, \apj, 500,
525

\bibitem[Sch\"odel et al.(2002)]{Sch1} Sch\"odel, R., Ott, T., Genzel, R., et al.  2002, \nat, 419,
694

\bibitem[Sch\"odel et al.(2003)]{Sch2} Sch\"odel, R., Ott, T., Genzel, R., et al.  2003, \apj, 596,
1015

\bibitem[Schwarzschild(1979)]{Sch79} Schwarzschild, M., 1979, \apj, 232,
236

\bibitem[Tremaine(1995)]{Tremaine} Tremaine, S.,  1995, \aj, 110,
628

\bibitem[Virbhadra \& Ellis(2000)]{VirEll} Virbhadra, K.S., \& Ellis, G.F.R. 2000, \prd, 62, 084003

\bibitem[Wardle \& Yusuf-Zadeh(1992)]{WarYus} Wardle, M., \& Yusuf-Zadeh, F. 1992, \apj, 387,
L65

\bibitem[Weinberg et al.(2005)]{Weinberg} Weinberg, N.N., Milosavljevic, M., \&
Ghez, A.M. 2005, ASP Conf. Proc. 338, 252

\bibitem[Widrow \& Dubinski(2005)]{WidDub} Widrow, L.M., \& Dubinski, J., 2005, \apj, 631,
838

\bibitem[Williams(2002)]{Wil} Williams, B.F., 2002, \mnras, 331,
293

\end{thebibliography}

\end{document}